\documentclass{emulateapj}
\usepackage{multirow}
\usepackage{xcolor}
\usepackage[T1]{fontenc}

\begin{document}

\def\be{\begin{eqnarray}}
\def\ee{\end{eqnarray}}
\def\bh{{\bullet}}
\def\edd{{\rm Edd}}
\def\in{{\rm in}}
\def\mb{{\rm mb}}
\def\ms{{\rm ms}}
\def\sal{{\rm Sal}}
\def\s{{\rm S}}
\def\d{{\rm d}}
\def\J{{\bf J}}
\def\disc{{\rm disc}}
\def\obs{{\rm obs}}
\def\c{{\rm c}}
\def\bol{{\rm bol}}
\def\ks{{\rm KS}}
\def\ini{{\rm ini}}
\def\fin{{\rm fin}}
\def\cl{{\rm cl}}
\def\al{{\rm al}}
\def\crit{{\rm crit}}
\def\sg{{\rm sg}}
\def\wp{{\rm warp}}
\def\bp{{\rm BP}}
\def\yr{{\rm yr}}
\def\acc{{\rm acc}}
\def\max{{\rm max}}
\def\foe{\left(\frac{f_\edd}{\eta_{0.1}}\right)}

\shorttitle{Constraining MBH growth history via spin
distribution}

\title{On Constraining the Growth History of Massive Black Holes via
Their Distribution on the Spin-Mass plane}
\author{Xiaoxia Zhang$^{1,2,3}$ and Youjun Lu$^{1,2,\dagger}$}
\affil{
~$^1$~National Astronomical Observatories, Chinese Academy of
Sciences, Beijing 100101, China;~$^{\dagger}$luyj@nao.cas.cn\\
~$^2$~School of Astronomy and Space Sciences, University of Chinese
Academy of Sciences, No. 19A Yuquan Road, Beijing 100049, China 
~$^3$~Department of Astronomy, Xiamen University, Xiamen, Fujian,
361005, China } 
\begin{abstract}
The spin distribution of massive black holes (MBHs) contains rich
information on the MBH growth history. In this paper, we investigate
the spin evolution of MBHs by assuming that each MBH experiences
two-phase accretion, with an initial phase of coherent-accretion via
either the standard thin disc or super-Eddington disc, followed by a
chaotic-accretion phase composed of many episodes with different disc
orientations. If the chaotic-phase is significant to the growth of an
MBH, the MBH spin quickly reaches the maximum value because of the
initial coherent-accretion, then changes to a quasi-equilibrium state
and fluctuates around a value mainly determined by the mean ratio of
the disc to the MBH mass ($M_\bullet$) in the chaotic-accretion
episodes, and further declines due to late chaotic-accretion if
$M_\bullet\ga(1-3)\times10^8M_\odot$.  The turning point to this
decline is determined by the equality of the disc warp radius and disc
size.  By matching the currently available spin measurements with mock
samples generated from the two-phase model(s) on the spin-mass plane,
we find that MBHs must experience significant chaotic-accretion phase
with many episodes and the mass accreted in each episode is roughly
1-2 percent of $M_\bullet$ or less.  MBHs with
$M_\bullet\ga10^8M_\odot$ appear to have intermediate-to-high spins
($\sim0.5-1$), while lighter MBHs have higher spins ($\ga0.8$). The
best matches also infer that (1)\,the radiative efficiencies ($\eta$)
of those active MBHs appear to slightly decrease with $M_\bullet$;
however, the correlation between $\eta$ and $M_\bullet$, if any, is
weak; (2)\,the mean radiative efficiency of active MBHs is
$\left<\eta\right>\sim0.09-0.15$, consistent with the global
constraints.
\end{abstract}

\keywords{accretion, accretion discs; black hole physics; galaxies:
active; galaxies: nuclei; relativistic processes  }

\section {Introduction}
\label{sec:Intro}

Observations have shown that massive black holes (MBHs) ubiquitously
exist in the centers of ellipticals and spiral bulges
\citep[e.g.,][]{kor95, mag98, kor13}. These MBHs are thought to be
completely described by the Kerr metric with only two physical
parameters, i.e., mass $M_{\bullet}$ and dimensionless spin parameter
$a$.  Measuring the masses and spins of these MBHs and obtaining their
distributions are of great importance in revealing their formation and
assembly histories.

The masses of MBHs in the centers of both nearby quiescent and active
galaxies can be estimated with considerable accuracy by using the
motion of gas and/or stars surrounding the MBHs \citep[e.g.,][]{mac97,
geb09, pet04}. Tight correlations have been found between the MBH mass
and properties of their host galaxies (e.g., stellar velocity
dispersions, bulge luminosities or masses, etc.) in the local universe
\citep[e.g.,][]{fer00, geb00, tre02, kor95, mag98, gul09, kor13,
sag16}, which suggests that MBHs co-evolve with their host galaxies.
It has also been shown that the mass growth of MBHs is dominated by
accretion during the QSO/AGN phases by comparing the local MBH mass
density with the accreted MBH mass density over the cosmic time
\citep[e.g.,][]{yu02, Marconi04, YL04, Shankar09} via the
\citet{1982MNRAS.200..115S} argument.

The spins of active MBHs are much more difficult to measure. Currently
it is widely agreed that they can be measured via the X-ray reflection
spectroscopy based on the assumption that the accretion disc is
geometrically thin and optically thick, and the inner disc boundary,
i.e., the innermost stable circular orbit (ISCO), is solely determined
by the MBH spin (\citealt{BR06}; for a review of MBH spin
measurements, see \citealt{bre13, rey14}). The most significant
feature in the X-ray reflection spectrum is the relativistically
broadened and skewed Fe K$\alpha$ line \citep[e.g.,][]{fab89, lao91},
the profile of which provides a measure to the spin of the central MBH
\citep[e.g.,][]{Tanaka95}. However, X-ray spectroscopic observations
of QSOs/AGNs with sufficiently high quality are very limited.
Currently, there are only about two dozen MBHs that have relatively
robust spin measurements, as summarized in Table~\ref{tbl-1}
\citep[see also][]{rey14, bre13, vas16}.  If ignoring the measurement
errors, all MBHs in this sample have spins $a>0.4$, and about two
thirds of them have $a>0.8$.

If an MBH merges with another MBH or accretes gaseous material, the
spin of the MBH evolves. Mergers of two MBHs with comparable mass
generally result in a spin value of $\sim 0.7 - 0.9$
\citep[e.g.,][]{Gammie04, Centrella10, Lousto10, Lehner14}, but may
leave little long-term effect on the MBH spin evolution
\citep{King08}. However, the MBH spin may increase or decrease by
accreting gaseous material, depending on the relative orientation of
the accretion disc angular momentum with respect to the MBH spin
\citep[e.g.,][]{kin06, per09}, although the MBH mass grows
monotonically. Significant gas accretion plays a critical role in the
evolution of MBH spin and may lead to a spin distribution over a large
range (e.g., \citealt{King08, Volonteri13, dot13}). It has been shown
that an MBH may end up with an extremely high spin (close to 1) if the
disc accretion has a preferred direction (hereafter
coherent-accretion; e.g., \citealt{tho74}). However, the MBH may be
spun down or up if the accretion is episodic with disc orientations
randomly distributed in different episodes (hereafter chaotic
accretion), and the MBH can be spun down to $a\sim 0$ even if the
initial spin is close to $1$ simply because negative angular momentum
injection by retrograde accretion is more effective than positive
angular momentum injection by prograde accretion
\citep[e.g.,][]{bar72,Moderski98}. In principle, different accretion
and assembly histories of MBHs may result in different spin
distributions, and thus observationally determined spin distribution
can be used to put constraints on the MBH accretion history
\citep[e.g.,][]{dot13, ses14, li15}.

In this paper, we use the latest available spin measurements for more
than two dozen MBHs to put constraints on the accretion history of
MBHs and check whether chaotic accretion is important in growing MBHs
and shaping the spin distribution.  The paper is organized as follows.
In Section~\ref{sec:spinevol}, we describe the spin evolution for
different accretion modes, including coherent super-Eddington
accretion, coherent standard thin disc accretion, and chaotic standard
thin disc accretion. We assume that MBHs experience an initial
coherent-accretion phase with either a super-Eddington rate or a
standard thin disc rate, followed by a chaotic-accretion phase with
each episode via standard thin disc accretion. Assuming this two-phase
accretion model, we use the currently available spin measurements to
obtain constraints on the accretion history of MBHs in
Section~\ref{sec:constraint}. Some discussions are given in
Section~\ref{sec:discussion}, and conclusions are summarized in
Section~\ref{sec:summary}. 
 
\begin{table*}
\begin{center}
\caption{Summary of the published MBH spin measurements via X-ray 
reflection spectroscopy}
\label{tbl-1}
\begin{tabular}{ccccccc}
\tableline
\tableline
Object  name 	& Galaxy type   & z          & $\log \left(L_\bol/
{\rm erg\,s^{-1}}\right)$      &  $M_{\bullet} (10^6 M_\sun)$     & a 	& Mass/Spin Refs\\
%
\tableline
1H 0419-577 	& \---- 	& 0.1040 & 46.03  & $\sim340$ 	   & $>0.89$          & ZW05/Wa13     \\
1H 0707-495 	& \---- 	& 0.0407 & 44.43  & $\sim2.3$  	   & $>0.97$         & ZW05/Zo10      \\ 
3C 120 	       	& S0		& 0.0330 & 45.34  &$55^{+31}_{-23}$& $>0.95$         &  Pe04/Lo13      \\ 
Ark 120		& Sb/pec	& 0.0327 & 44.91  & $150\pm19$     & $0.64^{+0.19}_{-0.11}$	&  Pe04/Wa13  \\
Ark 564		& SB		& 0.0247 & 44.21  & $\sim1.1$ 	   & $0.96^{+0.01}_{-0.07}$	& ZW05/Wa13  \\
Fairall 9	& Sc		& 0.0470 & 45.23  & $255\pm56$     & $0.52^{+0.19}_{-0.15}$	&  Pe04/Lo12  \\
H 1821+643	& \----		& 0.2970 & 47.30  & $4500\pm1500$  & $>0.4$      & Re14/Re14          \\ 
IRAS 00521-7054	& \----		& 0.0689 & \----  &  \----	   & $\ge 0.84$        & \----/Ta12    \\ 
IRAS 13224-3809	& \----		& 0.0658 & 45.55  & $\sim6.3$  	   & $>0.985$ 	& GS12/Fa13       \\
MCG 6-30-15	& E/S0		& 0.0570 & 44.18  &$2.9^{+1.8}_{-1.6}$& $>0.98$         & Mc05/BR06      \\
Mrk 1018	& S0		& 0.0424 & 44.39  & $\sim141$  	   & $0.58^{+0.36}_{-0.74}$	& Be11/Wa13 \\
Mrk 110		& \---- 	& 0.0353 & 44.71  & $25.1\pm6.1$   & $0.96^{+0.03}_{-0.07}$	& Pe04/Wa13 \\
Mrk 335		& S0a		& 0.0258 & 44.69  & $14.2\pm3.7$   & $>0.91$          & Pe04/Ga15      \\
Mrk 359		& pec		& 0.0174 & 43.55  & $\sim1.1$      & $0.66^{+0.30}_{-0.54}$	& ZW05/Wa13 \\
Mrk 79		& SBb		& 0.0222 & 44.57  & $52.4\pm14.4$  & $0.7\pm0.1$     & Pe04/Ga11        \\
Mrk 841		& E		& 0.0364 & 45.64  & $\sim79$  	   & $>0.52$	   & ZW05/Wa13    \\
NGC 1365	& SB(s)b	& 0.0055 & 43.48  & $\sim2$  	   & $\ge 0.84$	     & Ri09/Ri13  \\ 
NGC 3783	& SB(r)ab	& 0.0097 & 44.41  & $29.8\pm5.4$   & $\ge 0.98$	 & Pe04/Br11       \\
NGC 4051	& SAB(rs)bc	& 0.0023 & 43.56  & $1.91\pm0.78$  & $>0.99$     & Pe04/Pa12           \\
NGC 4151	& SAB(rs)ab	& 0.0033 & 43.73  & $45.7^{+5.7}_{-4.7}$& $>0.9$     & Be06/Ke15           \\
NGC 5506	& Sa		& 0.0062 &$\sim44$& $5.11^{+2.20}_{-1.18}$ & $0.93\pm0.04$		& Ni09/Su17 \\
Q 2237+305 	&\----		 &1.695	  &$\sim45$ &$1200\pm1080$      & $0.74^{+0.06}_{-0.03}$ &  Ass11/Rey14	\\ 
RBS 1124	& \----		& 0.2080 & 45.53  & $\sim180$ 	   & $>0.97$	    & Mi10/Wa13   \\
RXS J1131-1231 & \----		& 0.658  &$\sim45$& $\sim200$ 	   & $0.87^{+0.08}_{-0.15}$	& Sl12/Rei14	\\
SDSS J094533.99+100950.1 &  \----	& 1.66   & 46.79  & $\sim 2700$	   & $0.8^{+0.2}_{-0.5}$	& Cz11/Cz11 \\
Swift J0501.9-3239	& SB0/a(s)/pec	& 0.0124 & 44.11  & $45\pm15$ 	   & $\ge0.92$       & Ag14/Wa13        \\ 
Swift J2127.4+5654& \----	& 0.0144 & 44.53  &  $\sim15$	   & $0.6\pm0.2$       & Ma08/Mi09    \\
Ton S180	& \----		& 0.0620 & 45.30  & $\sim8.1$      & $0.92^{+0.03}_{-0.11}$	& ZW05/Wa13 \\
\tableline
\end{tabular}
\end{center}
\tablecomments{ 
The columns from left to right represent (1) Object name; (2) Galaxy type; (3) Redshift; 
(4) Logarithmic bolometric luminosity; (5) Mass in unit of $10^6 M_\odot$; (6) Spin; 
(7) References for the MBH mass/spin measurement. The masses are quoted with
1$\sigma$ errors, while the spins are quoted with 90\% confidence level. 
The references for the MBH masses are as follows: ZW05=\citet{zhou05}; 
Pe04=\citet{pet04}; Re14=\citet{reynolds14}; GS12=\citet{gs12}; 
Mc05=\citet{mchardy05}; Be11=\citet{bennert11}; Ri09=\citet{risaliti09}; Be06=\citet{bentz06}; 
Ni09=\citet{nikolajuk09};  
Ass11=\citet{assef11}; Mi10=\citet{miniutti10}; 
Sl12=\citet{sluse12}; Cz11=\citet{czerny11}; Ag14=\citet{ag14}; Ma08=\citet{malizia08}.
The references for the spins are as listed:
Wa13=\citet{walton13}; Zo10=\citet{zoghbi10}; Lo13=\citet{lohfink13}; Lo12=\citet{lohfink12}; 
Re14=\citet{reynolds14}; Ta12=\citet{tan12}; Fa13=\citet{fabian13}; 
BR06=\citet{BR06}; Ga15=\citet{gallo15}; Ga11=\citet{gallo11}; Ri13=\citet{risaliti13}; 
Br11=\citet{brenneman11}; Pa12=\citet{patrick12}; Ke15=\citet{keck15}; Su17=\citet{sun17}; 
Rey14=\citet{rey_mt14}; Rei14=\citet{Reis14};
Cz11=\citet{czerny11}; Mi09=\citet{miniutti09}. Note here that IRAS 00521-7054 has no 
mass measurement, and it is not used in later model constraints. 
}
\end{table*}

\section{Spin evolution of MBHs}
\label{sec:spinevol}

\subsection{Accretion history of MBHs}
\label{subsec:acchis}

If a galaxy is rich in gas, once it experiences violent perturbations such 
as major mergers or disc instabilities, a large fraction of gas will be poured 
into the galactic center \citep[e.g.,][]{Krolik99}, which 
naturally triggers disc accretion (either sub-Eddington or 
super-Eddington) onto the central MBH. With the consumption of  
gaseous material, the accretion process may be episodic afterwards due 
to infalling of single gas cloud, and in different episodes the disc angular 
momentum could be randomly oriented with respect to the MBH spin 
if there is no mechanism to make the gas cloud infall with a preferred direction.

As suggested by demographic studies of SDSS QSOs and X-ray AGNs
\citep[e.g.,][]{she08, sch10, Suh15}, most QSOs and AGNs are accreting
via sub-Eddington rate, i.e., $L_{\rm bol}/L_\edd \sim 0.01 - 1$. Here
$L_{\bol}$ is the bolometric luminosity and $L_\edd=\frac{4\pi G M_\bh
m_{\rm p} c}{\sigma_{\rm T}} \simeq 1.3 \times 10^{46} \left(
\frac{M_\bh}{10^8 M_\odot} \right) {\rm erg\, s^{-1}}$ is the
Eddington luminosity, with $G$ the gravitational constant, $m_{\rm p}$
the proton mass, $c$ the speed of light and $\sigma_{\rm T}$ the
Thomson cross section. With such a moderate accretion rate, the disc
is radiatively efficient and can be described by the standard thin
disc model \citep{sha73, NT73}. However, there are two lines of
observations suggesting that MBHs may accrete material via
super-Eddington rate in the early stage of its growth.  First, MBHs
with mass $> 10^9 M_\odot$ have already formed at redshift $z>6$ when
the universe is less than $1$\,Gyr old \citep[e.g.,][]{mor11, wu15,
Banados18}. These observations raise a significant challenge to the
growth theory for those MBHs since an MBH with mass $>10^9M_\odot$
cannot grow up from a small seed black hole (e.g., $<100M_\odot$) via
the Eddington-limited accretion within a time period $<1$\,Gyr. One
popular solution to this is that those MBHs accrete via a
super-Eddington rate, at least at the early stage
\citep[e.g.,][]{Li12, mad14}.  Second, a number of nearby AGNs are
recently found to be accreting via super-Eddington rate by adopting
the MBH masses estimated from the reverberation mapping technique
\citep[e.g.,][]{Du15}. For such super-Eddington accretion flows, thick
discs will be formed around the central MBHs \citep[e.g.,][]{abr88},
as the emitted photons are trapped by the high density accretion flows
and the discs cannot be  cooled efficiently.  

In addition, it has been shown that the net lifetime of QSOs is larger
than $10^7-10^8$\,yr \citep[e.g.,][]{yu08, Shankar09}, while the
period of a single accretion epoch could be as short as
$10^4-10^6$\,yr \citep[e.g.,][]{Martini04}, which suggests the
accretion processes may be episodic and the time duration for
individual episode can be substantially shorter than the net lifetime.
Recent observations also reveal a number of changing-look AGNs on
timescale of $\sim 10$\,yr, which are probably due to significant
changes in the accretion rate \citep[e.g.,][]{LaM15, Ruan16}. These
may also suggest that MBH accretion is episodic, especially at its
later growth stage.

According to the above observational results, we assume a two-phase
accretion model to describe the growth of MBHs. In the first phase,
MBHs experience continuous and coherent accretion, during which the
disc orientation maintains the same. In this phase, the accretion rate
could be below or above the Eddington rate. After that, those MBHs
undergo chaotic thin disc accretion, which contains many accretion
episodes, and in each episode the disc angular momentum is arbitrarily
oriented with respect to the MBH spin. We ignore the MBH change due to
mechanisms in between any two adjacent accretion episodes.

We define normalized accretion rate as $\dot{m}=\dot{M}/\dot{M}_\edd$,
where $\dot{M}_\edd \simeq 16\ L_\edd/c^2$  (according to
\citealt{mad14}) is the critical accretion rate for a non-rotating MBH
whose accretion disc radiates at Eddington luminosity $L_\edd$. The
accretion is super-Eddington if $\dot{m} > 1$.

\subsection{Spin evolution due to coherent and chaotic accretion}

Regardless of accretion patterns, if an MBH accretes with Eddington
ratio $f_\edd$ and radiative efficiency $\eta$, the MBH growth rate
can be expressed as
\be
\frac{\d M_\bh}{\d t}=(1-\eta) \frac{f_\edd}{\eta}
\frac{M_\bh}{t_\edd},
\label{eq:dmdt}
\ee where $t_\edd \equiv M_\bullet c^2/L_\edd \simeq 4.5\times
10^8$\,yr is the Eddington timescale. The above equation is obtained
by assuming that the kinetic energy loss is negligible and
$\eta=1-E(R_{\rm in})$, where $E(R_{\rm in})$ 
is the specific energy at the inner disc boundary $R_{\rm
in}$.

Generally, the initial disc angular momentum can be misaligned with
the MBH spin, and the disc will suffer from warps due to the
Lense-Thirring (LT) precession \citep{len18}. Since the LT precession
frequency decreases with increasing radius ($\propto R^{-3}$), the
inner disc may be bent to the MBH equatorial plane while the outer
disc maintains the original orientation with a transiting warped
region in between \citep{bar75}.  In such a case, the evolution of the
MBH spin vector ($\J_\bh$) is governed by \citep[e.g.,][]{lodato06,
per09} 
\be
\frac{\d\J_\bh}{\d t} = \dot{M} \frac{G M_\bh}{c}\Phi(R_{\rm
in})\hat{\bf l}+\frac{4\pi G}{c^2}\int_\disc\frac{{\bf L}\times
\J_\bh}{R^2} \d R, \nonumber \\
\label{eq:djdt}
\ee 
where 
$\Phi$ is the specific angular momentum of the accreted material
at the disc inner boundary, $\hat{\bf l}$ is a unit vector
parallel to $\J_\bh$, and ${\bf L}$ is the angular momentum of
per-unit-area disc. The first term on the r.h.s of
Equation~(\ref{eq:djdt}) only leads to the modification of the spin
modulus, while the second term, dominated by the contribution from
outer disc, describes the gravito-magnetic interaction between the
disc and MBH, and only causes the variation of the spin direction.
We note here the absolute spin parameter is defined as $|a|\equiv
cJ_\bullet/GM_\bullet^2$ with $J_\bullet=|\J_\bh|$, and $a$ is
positive if the disc is co-rotating around the MBH and negative if
otherwise. The canonical value $0.998$ is set as the upper limit of
the spin, i.e., $-0.998 \le a \le 0.998$.  

Equations~(\ref{eq:dmdt}) and (\ref{eq:djdt}) are general formulas
governing the mass and spin evolution of MBHs under accretion.
However, for different accretion modes, the quantities involved could
be different, which are discussed separately as follows.\footnote{
Note that we set the canonical value of 0.998 \citep{tho74} as an
upper limit of the MBH spin throughout the paper, and we also ignore
the photon trapping effect since it is important only when $a>0.99$
\citep{tho74}, which is not the focus of this paper.  }

\begin{figure}
\centering
\includegraphics[width=3.4in]{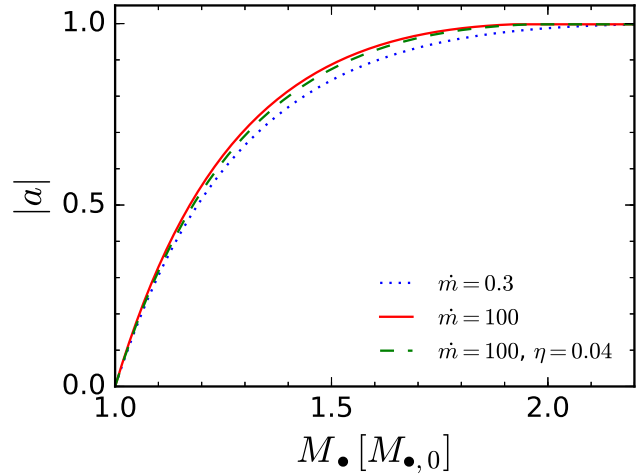}
\caption{MBH spin evolution as a function of mass (in unit of the
initial mass $M_{\bh, 0}$) due to coherent-accretion. The blue dotted
(red solid) line represents the case of the thin (thick) disc
accretion with a constant rate $\dot{m}=0.3\ (100)$ and radiative
efficiency $\eta$ derived from $f_\edd-\ln\dot{m}$ relation; the green
dashed line represents the thick disc accretion with a constant rate
of $\dot{m}=100$ and $\eta = 0.04$ \citep[for this efficiency,
see][]{jia14}. 
}
\label{fig:f1}
\end{figure}

{\bf (i) Coherent thin disc accretion.} For continuous and coherent
accretion with a moderate rate $\dot{m} \lesssim 1$, the disc is
assumed to be described by the standard thin disc model. Then the
inner boundary of the disc is the ISCO, which is solely determined by
the MBH spin, and the specific energy $E$ and angular momentum $\Phi$
at $R_{\rm ISCO}$ can be obtained \citep [see Appendix~\ref{sec:app1}
for expressions of $R_{\rm ISCO}$ as a function of spin and dependence
of $E$ and $\Phi$ on radius for given spins; e.g.,][]{bar72}.  For the
coherent case considered here, the alignment timescale is much shorter
than the accretion (or viscous) timescale. Therefore, we ignore the
initial short time period for the alignment and assume the MBH spin is
instantaneously aligned with the total angular momentum of the system,
which is dominated by the disc. In this case, the second term on the
r.h.s. of Equation~(\ref{eq:djdt}) vanishes. Combining
Equations~(\ref{eq:dmdt}) and (\ref{eq:djdt}), we derive the following
equation that governs the spin modulus evolution \be
\frac{\d a}{\d t}  &=& [\Phi(R_{\rm in})-2a \ (1-\eta)] \
\frac{f_\edd}{\eta \ t_\edd}, 
\label{eq:dadt}
\ee where $R_{\rm in}=R_{\rm ISCO}$. An example for the spin evolution
is shown in Figure~\ref{fig:f1} (dotted line), obtained for the
coherent thin disc accretion case by assuming $\dot{m}=0.3$.

{\bf (ii) Coherent super-Eddington accretion.} For accretion flows
with infalling rate of $\dot{m}>1$, the heat produced in the disc
cannot be released efficiently, resulting in an inflated (or a thick)
disc. For the mass evolution [Eq.~(\ref{eq:dmdt})], the specific
energy (and efficiency $\eta$) at the inner boundary $R_{\rm in}$ of a
thick disc is different from that of a thin disc. It is believed that
even if the accretion rate is highly super-Eddington, the disc
luminosity can only mildly exceed the Eddington limit, leading to a
commonly adopted assumption that $f_\edd$ logarithmically depends on
$\dot{m}$ at $\dot{m}>25/8$ \citep[the $f_\edd-\ln\dot{m}$ relation;
Eq.~(\ref{eq:fedd}); e.g.,][]{min00}.  For the thick disc accretion,
$R_\in$ is in between the ISCO and marginally bound orbit
\citep{koz78, Jaro80}, and we can obtain $R_\in$ by interpolation (see
Appendix~\ref{sec:app2}). Similarly, we ignore the short alignment
timescale and solve Equation~(\ref{eq:dadt}) to obtain the spin
evolution.

The spin evolution for an MBH accreting via $\dot{m}=100$ (0.3) is
shown as the red solid (blue dotted) line in Figure~\ref{fig:f1},
where the MBH becomes maximally spinning when its mass roughly
doubles.  Note that our result for $\dot{m}=100$ is only slightly
different from that given by \citet{sad11}, in which a detailed slim
disc accretion model is adopted to solve the spin evolution equation,
with the consideration of photon capture effect. They found that the
spin evolution depends on both the viscosity $\alpha$ and the
accretion rate $\dot{m}$, and the maximum spin value $a_\max$ is
slightly different from the canonical value $0.998$. For example, with
$\alpha=0.1$ and $\dot{m}=1$, the MBH reaches a maximum spin value of
$0.9924$ when its mass becomes $\sim2.4$ times larger.  We do not
repeat the complicated calculations by \citet{sad11} for thick disc
accretion, as the slight difference in the maximum spin value and the
time it reaches the value does not affect our final results much. 

Numerical solution of the relativistic slim disc equations
\citep{sad09} found that $f_\edd$ is also dependent on the MBH spin.
\citet{mad14} fit the $f_\edd-\dot{m}$ relation by adding additional
dependence on spin according to the simulation results of
\citet{sad09}. We have checked that adopting this spin-dependent
relation makes little difference to the spin evolution.

\citet{jia14} claim that super-Eddington accretion could be
radiatively more efficient than the $f_\edd-\ln\dot{m}$ relation
gives. Using magneto-hydrodynamic simulations, they found
$\eta\sim0.04$ for super-Eddington accretion by considering the
buoyancy effect. We further check the spin evolution in this case by
setting constant efficiency $\eta=0.04$, accretion rate $\dot{m} =
100$, and the Eddington ratio $f_\edd=16\eta \dot{m}$ (the green
dashed line in Fig.~\ref{fig:f1}). It appears that the resulting spin
evolution shows little difference comparing to the case of either thin
disc accretion with a rate of $\dot{m}= 0.3$ (blue dotted line in
Fig.~\ref{fig:f1}) or thick disc accretion with $\dot{m}=100$ (red
solid line in Fig.~\ref{fig:f1}), both with $\eta$ inferred from the
$f_\edd-\ln\dot{m}$ relation.

For the coherent-accretion phase, we conclude that different choices
of the accretion rate result in only slight difference in the spin
evolution as a function of mass (see Fig.~\ref{fig:f1}).

{\bf (iii) Chaotic thin disc accretion.} 
We consider multi-episode accretion of gas clouds, and in each
episode the clouds infall with random orientations. The accretion rate
is assumed to be moderate, i.e., $f_\edd=0.3$, and in this case the
disc is described by the standard thin disc model (but not necessarily
on the MBH's equatorial plane).  Choosing a different $f_\edd$ mainly
affects the accretion time, and makes little difference to the
spin-mass evolutionary curves.  The disc orientation is given by the
polar angle $\theta$ and azimuthal angle $\phi$ relative to $Oxyz$
(the observer's rest frame centered on the MBH with $z$ as the
direction from the MBH to the distant observer).  For each chaotic
episode, $\phi$ and $\theta$ are randomly selected from a flat
probability distribution over $0$ to $2\pi$ and a probability
distribution proportional to $\sin\theta$, respectively, in order to
achieve random orientations of the chaotic discs.  Different from the
coherent thin disc case, the discs considered here are much smaller, and
the temporal evolution of the disc involving the LT effect has to be
considered.  For this part, the procedures to calculate the spin
evolution are similar to that provided by \citet[][their model with
$F=0.5$]{dot13}, and the only difference is the disc mass in each
episode.

The amount of gas that is available for accretion is probably not the
same in different accretion episodes. We assume that in each episode,
the mass of the gas cloud infalling and to be accreted depends on the
MBH mass and is described by \be 
M_\cl= b  M_\bh \left( \frac{M_\bh}{10^8 M_\odot}\right)^{\gamma},
\label{eq:mcl} \ee where $b$ and $\gamma$ are constant parameters,
$M_\bh$ is the MBH mass at the beginning of each episode. If
$\gamma=0$, then $M_\cl$ scales linearly with $M_\bh$; if $\gamma=-1$,
then $M_\cl$ is a constant and is irrelevant to the MBH mass but
determined by the environment. In this section, we aim at illustrating
how the MBH spin evolves for different settings of $M_\cl$, and only
consider the case with $\gamma=0$ for simplification. In
Section~\ref{sec:constraint}, we will consider more general cases with
$\gamma \neq 0$.

The whole cloud is assumed to form an accretion disk with
negligible mass loss, i.e., $M_\disc=M_\cl$. The disc size ($R_\disc$)
is  estimated via Equation~(\ref{eq:rbh}) by applying the surface
density profile of standard thin disc, and is then compared with the
warp radius ($R_\wp$; Eq.~\ref{eq:rwp}) of the disc which
approximately marks the distance of maximally warped region to the
central MBH. If $R_\disc>R_\wp$, then Equations~(\ref{eq:dmdt}) and
(\ref{eq:djdt}) are solved by applying the adiabatic approximation
\citep{per09}, i.e., the disc transits through a sequence of steady
warped states over a short time interval $\delta t \ll t_\al$, where
$t_\al$ is the alignment timescale.  The analytic solution of how the
disc is deformed at different radii has been found by \citet{mar07},
and the analytic expression of the torque term in
Equation~(\ref{eq:djdt}) with respect to the MBH coordinate ($Ox'y'z'$;
$z'$ is always parallel to $\J_\bh$) is directly provided by
\citet[][see their Appendix for power-law viscosity]{per09}.  The
variation of $\J_\bh$ within each $\delta t$ with respect to $Ox'y'z'$
is then rotated back to the observer's rest frame \citep[see][for
details]{per09, dot13}. 

If the inequality $R_\wp>R_\disc$ holds, then the angular momentum of
the MBH is assumed to be instantaneously aligned with the total
angular momentum $\J_{\rm tot} (=\J_\disc+\J_\bh)$. The re-orientation
of the MBH spin is rather small since $\J_\bh$ dominates over
$\J_\disc$.  The disc goes through a fast and significant
re-orientation, and whether the disc angular momentum is aligned or
anti-aligned with the MBH spin is determined by the disc-to-MBH
angular momentum ratio and the angle $\beta$ between ${{\bf J}_{\rm
disc}}$ and ${\bf J}_\bullet$. If $\cos\beta >-J_{\rm disc}
/2J_\bullet$, then they are aligned; otherwise, anti-aligned
\citep{kin05}. In this case, we only need to solve
Equation~(\ref{eq:dadt}) that governs the spin module evolution for
each single chaotic accretion episode. This is similar to that for the
coherent accretion case, except that anti-alignment is possible here
and the direction of the MBH spin is re-oriented to the total angular
momentum direction for each chaotic accretion episode.

It is proposed that the accretion disc cannot be too massive, as it
may be unstable against its own gravity and can be fragmented into gas
clumps at the outer region when the disc is too massive
\citep[e.g.,][]{kol80, goo04, King08}. The criterion for disc
instability is given by Toomre-$Q=1$, which yields a maximum disc size
$R_{\disc,\sg}$, and this corresponds to a maximum disc mass $M_\sg$
(see Appendix~\ref{sec:app3} for details).  Although there could be
such an upper limit for the disc mass, the infalling of gas clumps
onto the outer disc and other mechanisms may also heat the disc
significantly and thus prevent it from fragmentation. Therefore, we
consider two cases: one case is that the disc is not affected by the
possible instabilities due to its self-gravity (not limited by
$M_\sg$) for which $M_\disc=M_\cl$, and the other is that the disc
mass is indeed regulated by self-gravity, i.e., $M_\disc=\min(M_\cl,
M_\sg)$. If not otherwise stated, we will mainly focus on the former
case, while the latter case will be investigated in details in
Section~\ref{sec:constraint}.

\begin{figure}
\centering
\includegraphics[width=3.4in]{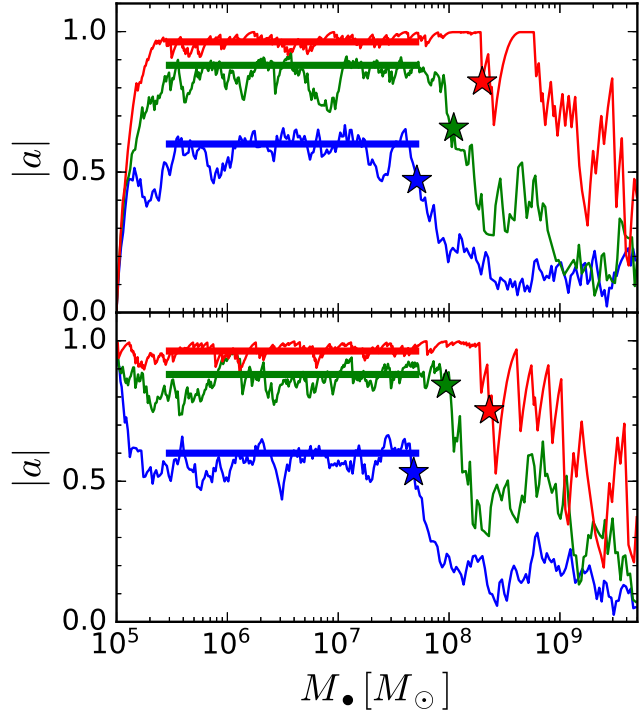}
\caption{Examples for spin modulus evolution of MBHs undergoing
chaotic accretion. These MBHs are assumed to have initial masses of
$10^5M_\odot$, initial spins of either $0$ (top panel) or $0.998$
(bottom panel), and grow up via chaotic accretion with many episodes
with $M_\disc=b M_\bh$ in each episode, where $b=0.003$ (blue line),
$0.01$ (green line), or $0.03$ (red line). Star symbols mark the
critical MBH mass given by $R_\disc=R_\wp$. For each curve shown here,
the horizontal line with the corresponding color indicates that the
spin modulus oscillates around an equilibrium value when the MBH mass
is smaller than the critical MBH mass.  }
\label{fig-mbh}
\end{figure}

\begin{figure}
\centering
\includegraphics[width=3.4in]{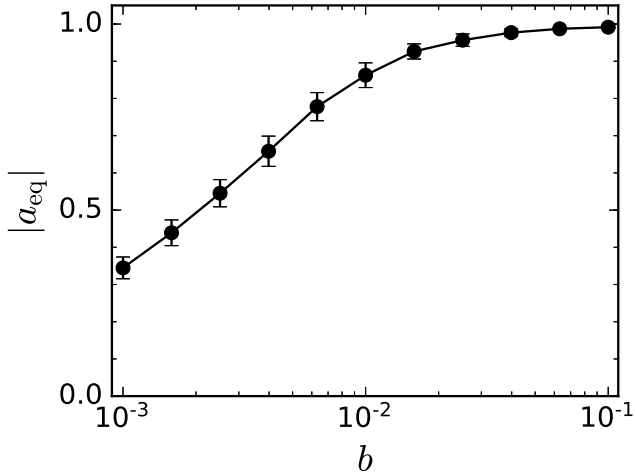}
\caption{Dependence of the equilibrium spin value on the disc mass
(characterized by $b$) for MBHs undergoing chaotic accretion (see the
horizontal lines shown in Fig.~\ref{fig-mbh}). The black circles and
their associated error bars show the mean values and their standard
deviations resulting from 200 realizations of Monte Carlo simulations.
}
\label{fig-ab}
\end{figure}

Figure~\ref{fig-mbh} shows examples of spin evolution for several MBHs
accreting chaotically with $M_\disc=b M_\bh$ in each episode. Whatever
the initial spin is ($0$ or $0.998$), the MBH spin quickly reaches a
quasi-equilibrium state when the MBH mass roughly doubles and then
fluctuates around the equilibrium value until $M_\bh \sim 10^8
M_\odot$.  For initially non-rotating MBHs, the sharp increase at the
beginning is due to the short alignment timescale. The MBH spin is
quickly re-aligned to  $\J_{\rm tot}$, which is approximately parallel
with $\J_\disc$  because $\J_\disc$ dominates over $\J_\bh$ when the
MBH mass is low, leading to an increase in spin with time. For
initially maximally spinning MBHs, the (sharp) decrease of the spin
can be similarly explained.  How large the equilibrium value $a_{\rm
eq}$ could reach depends on the disc mass in each episode, i.e., the
$b$ value. The dependence of $a_{\rm eq}$ on $b$ is shown in
Figure~\ref{fig-ab}, and each point with error bar (one standard
deviation) is obtained from $200$ realizations of Monte Carlo
simulations.\footnote{We note here that the dependence of $|a_{\rm
eq}|$ on $b$ is slightly affected by the settings of $\alpha$ and
$f_{\nu_2}$.} But in general choosing a different set of $\alpha$ and
$f_{\nu_2}$ only slightly affects those constraints obtained on MBH
growth obtained in Section~\ref{sec:constraint}.  How spin modulus
evolves is determined by the competition of prograde and retrograde
accretion. For ideal chaotic case, the number of prograde and
retrograde episodes are the same, and thus the MBH spin appears to
decrease with time since retrograde accretion is more efficient in
angular momentum injection. However, the spin also precesses and tends
to align with the disc angular momentum.  Hence, what matters is
whether the spin could align efficiently within a single accretion
episode. If yes, then the accretion will quickly transit to prograde
even if it starts with retrograde, and then the spin increases.

\begin{figure}
\centering
\includegraphics[width=3.4in]{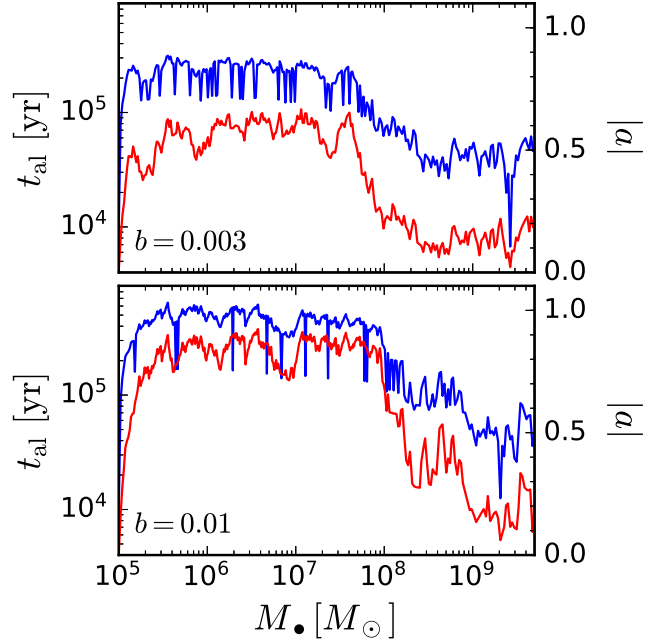}
\caption{Examples for the evolution of the alignment timescale $t_\al$
(left y-axis; blue lines) and the MBH spin modulus (right y-axis; red lines) 
with the mass growth of MBHs undergoing chaotic accretion with 
$M_\disc=0.003M_\bh$ (top panel) and $0.01M_\bh$ (bottom panel) 
in each episode. The red lines in the top (bottom) panel are the same 
as the blue (green) lines in the top panel of Fig.~\ref{fig-mbh}.
}
\label{fig-tal}
\end{figure}

Figure~\ref{fig-tal} shows coupled evolution of the alignment
timescale and the MBH spin for $b=0.003$ (top panel) and 0.01 (bottom
panel), where the alignment timescale is evaluated by $t_\al \approx
10^5\alpha^{58/35}_{0.1} f^{-5/7}_{\nu_2} M^{-2/35}_{\bh,6}
\left(\frac{f_\edd}{\eta_{0.1}}\right)^{-32/35} a^{5/7}  \rm{yr}$
\citep{per09}.  As seen from this figure (see also the blue and green
curve in Fig.~\ref{fig-mbh}), the spin roughly maintains at an
equilibrium value and oscillates around it when the MBH mass is around
a few times $10^5M_\odot$ to a few times $10^7M_\odot$. This
quasi-equilibrium state can be explained as follows. An increase in
spin leads to an increase in $t_\al$ ($\propto a^{5/7}$); a less
efficient alignment  leads to a decrease of spin; hence the spin
oscillates. With further increase of the MBH mass, the disc size may
reach a critical value of $R_\disc = R_{\rm warp}$, and after that the
MBH spin leaves the quasi-equilibrium state. When $R_\disc < R_{\rm
warp}$, the MBH and disc angular momenta are assumed to be
instantaneously aligned or anti-aligned with each other, and thus the
spin tends to decline on average with increasing MBH mass after
$R_\disc = R_{\rm warp}$ (the transition is shown as the star symbols
in Fig.~\ref{fig-mbh}). How fast the spin decreases is determined by
the relative fraction of prograde and retrograde episodes, which
mainly depends on $J_\disc/J_\bh$, and thus $b$ [see
Eq.~(\ref{eq:jdojh})]. 

\begin{figure}
\centering
\includegraphics[width=3.4in]{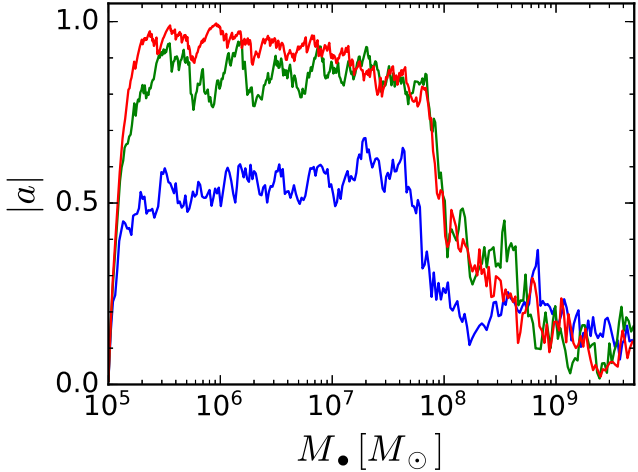}
\caption{Example spin modulus evolutionary curves for MBHs undergoing
chaotic accretion by considering the self-gravity of the disc in each
accretion episode. The MBH initial masses and spins are fixed to be
$10^5M_\odot$ and 0, respectively, and the disc mass in each accretion
episode is assumed to be $M_\disc=\min(b M_\bh, M_\sg)$, where
$b=0.003$ (blue), $0.01$ (green), or $0.03$ (red). The slight decline
of the equilibrium spin value with increasing $M_\bullet$ (from $\sim
3\times 10^5$ to $6\times 10^7M_\odot$) shown by the red line is due
to the decrease of the mean disc-to-MBH mass ratio .  }
\label{fig-sg}
\end{figure} 

\begin{figure*}
\centering
\includegraphics[width=7in]{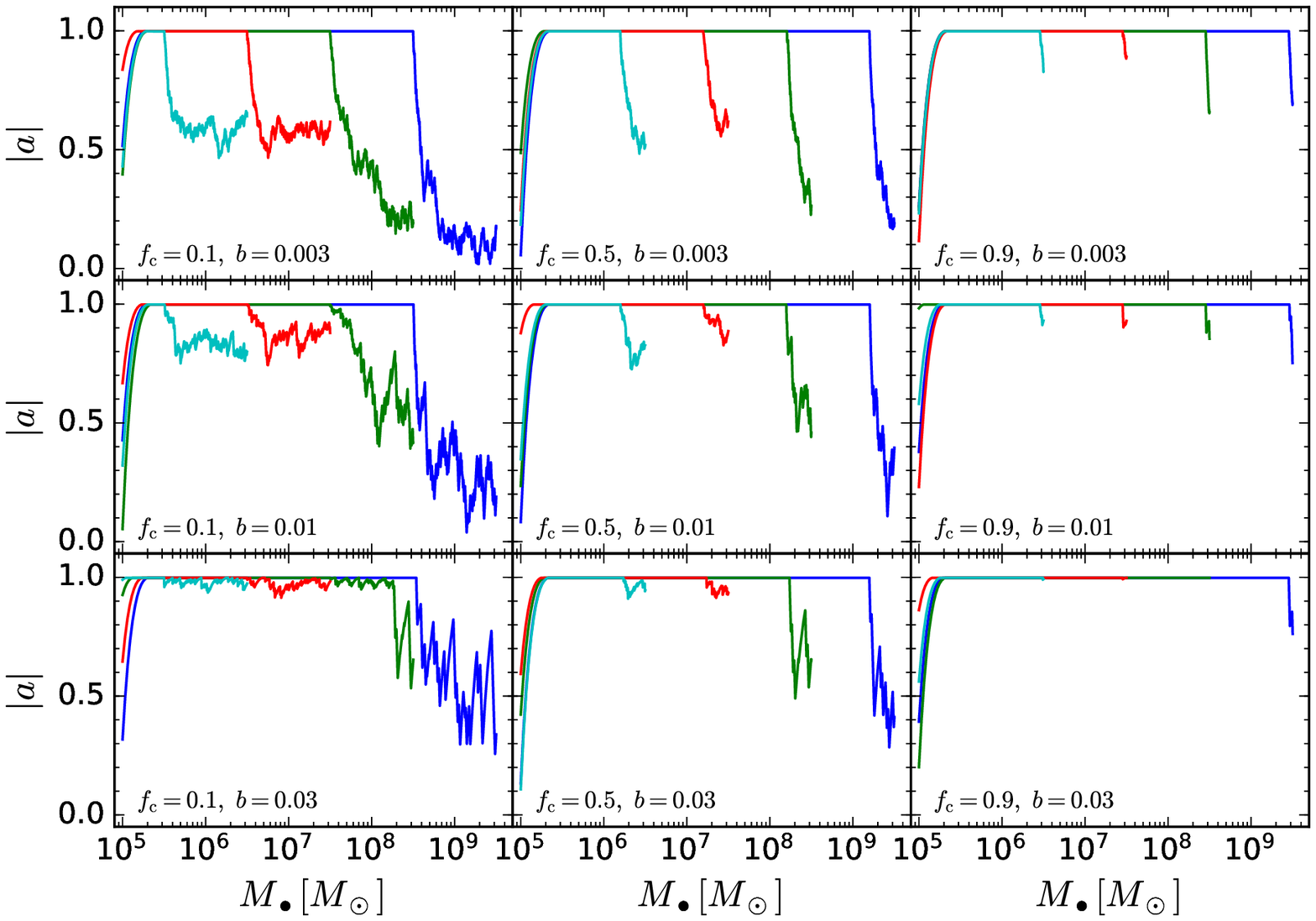}
\caption{Examples spin modulus evolutionary curves of MBHs undergoing
two-phase accretion with $\gamma=0$, $f_\c=0.1$ (left), $0.5$ (middle)
or $0.9$ (right), $b=0.003$ (top), $0.01$ (middle), or $0.03$
(bottom).  The final masses of the MBHs are $M_{\bullet,\rm f}=
3\times 10^6M_\odot$ (cyan), $3\times 10^7M_\odot$ (red), $3\times
10^8M_\odot$ (green), or $3\times 10^9M_\odot$ (blue).  The initial
masses here are fixed at $10^5 M_\odot$, and the initial spins are
randomly generated between 0 and 1.
}
\label{fig:f6}
\end{figure*}

Figure~\ref{fig-sg} shows the spin evolution of MBHs undergoing
chaotic accretion by considering the self-gravity of the accretion
disc. For the case with small $b$, i.e., $b=0.003$ (blue line), it is
similar to that shown in Figure~\ref{fig-mbh} without consideration of
disc self-gravity, because $M_\cl < M_\sg$ is valid almost in the
whole accretion history of the MBH. For $b=0.01$, however,
self-gravity plays an important role when $M_\bh \ga 2 \times10^7
M_\odot$ ($M_\cl \ga M_\sg$).  Therefore, the MBH spin magnitude
decreases faster and reaches a lower value when the MBH mass is high
(larger than a few times $10^8M_\odot$), compared with that shown in
Figure~\ref{fig-mbh}. For $b=0.03$, self-gravity is always important
in the accretion history of the MBH. In this case, the equilibrium
spin value slightly decreases with increasing MBH mass because of the
decrease of $b\ (=M_\sg/M_\bh)$ with increasing $M_\bh$, and the MBH
spin decreases as fast as that with $b=0.01$ when $R_\disc > R_{\rm
warp}$, and reaches similarly low values at the high-mass end (red
curve in Fig.~\ref{fig-sg}).

\section{Observational constraints on MBH accretion history via spin
and mass distribution}
\label{sec:constraint}

Currently there are more than two dozen AGNs that have relatively
robust spin measurements via the X-ray reflection spectroscopy (see
Table~\ref{tbl-1}) as mentioned in the Introduction. About two thirds
of these AGNs have spins $> 0.8$, and the rest have intermediate spins
$\sim  0.4 - 0.8$. In this section, we aim at using the distribution
on the mass versus spin plane of this sample to constrain the
accretion history of MBHs.

We assume that all MBHs experience an initial coherent-accretion phase
and a later chaotic-accretion phase. For an MBH with initial mass
$M_{\bullet,0}$ and final mass $M_{\bullet,\rm f}$, it grows up
through coherent accretion before its mass reaches a factor $f_{\rm
c}$ of $M_{\bullet,\rm f}$, and after that, it enters to the second
phase and grows via chaotic accretion as argued in
Section~\ref{sec:spinevol}. The disc mass in each chaotic accretion
episode is assumed to be described by
Equation~(\ref{eq:mcl}).
For such a two-phase accretion model, parameters that are interesting
in this paper are mostly $f_\c$ and $b$ if $\gamma$ is fixed.

\begin{figure*}
\centering
\includegraphics[width=7in]{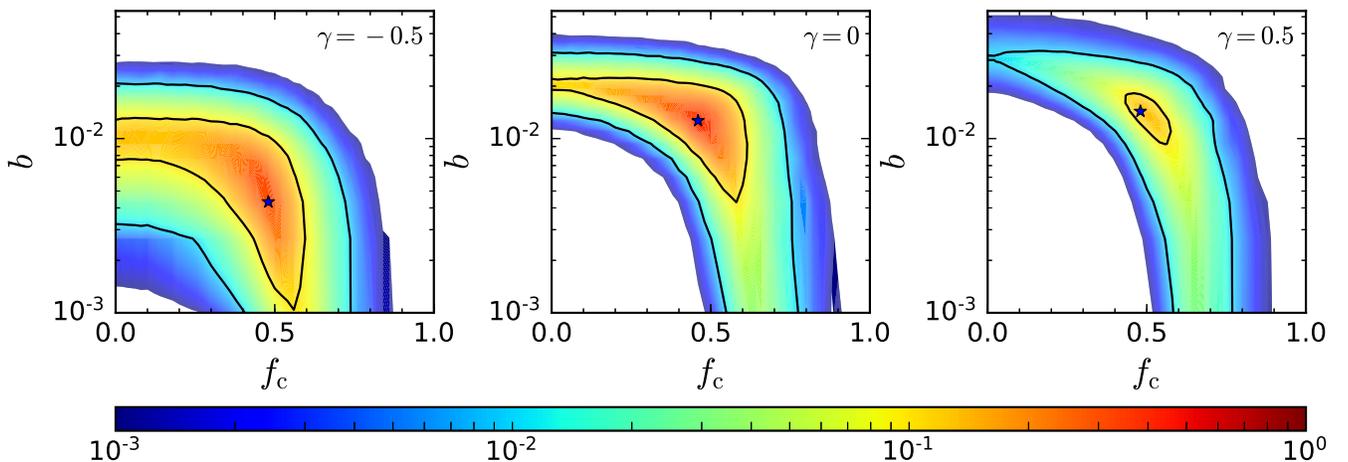}
\caption{Constraints on the parameters $b$ and $f_\c$ in the two-phase
accretion model from the available spin measurements of AGNs listed in
Table~\ref{tbl-1}. Colors represent the $P$-value obtained from the
2D-Kolmogorov-Smirnov (2D-KS) test as indicated by the color bar at
the bottom of this figure. Panels from left to right are for the cases
with $\gamma=-0.5$, $0$, and $0.5$, respectively. Star symbol in each
panel marks the location with the largest 2D-KS $P_\ks$ value, which
is $(f_\c,\ b) = (0.48,\ 0.004)$, $(0.46,\ 0.012)$, and $(0.48,\
0.014)$, respectively from the left to right panel. The black lines
mark $P_\ks=0.1$ and 0.01.  The white region that is not filled with
colors has $P_\ks<0.001$ and therefore can be excluded. Note here we
simply adopt the lower limit of the color bar as $0.001$ in order to
make the color gradient more clear for the rest parts.}
\label{fig:f7}
\end{figure*}

Figure~\ref{fig:f6} shows the spin evolutionary tracks of a few MBHs
by assuming such two-phase accretion models with $\gamma=0$ and
different sets of the other two parameters ($f_\c$, $b$).  If $f_c$ is
small and thus the chaotic-accretion phase dominates the MBH growth
(left panels of Fig.~\ref{fig:f6}), then the initial
coherent-accretion phase makes the spin increase quickly to the
canonical value $0.998$, and it maintains until the chaotic accretion
phase makes it decrease somewhat. Afterwards it oscillates around an
equilibrium value depending on the value of $b$. If $M_{\bullet,\rm
f}$ is larger than several times $10^8M_\odot$, the MBH spin further
decreases to small values (close to $0$) in the late chaotic accretion
phase. If $f_\c$ is large (e.g., $0.9$; right panels of
Fig.~\ref{fig:f6}), the chaotic-accretion phase is not significant for
the MBH growth and thus the de-spin may only occur in a short period
of the later life of a QSO since $f_\c$ determines fraction of its
lifetime with spin close to $1$ and the time left to the
chaotic-accretion phase (cf., right panels of Fig.~\ref{fig:f6}). Note
that the value of $b$ (or $M_\disc$ in each episode) determines how
efficiently the MBH could be spun down in a single chaotic accretion
episode.

In order to obtain constraints on the MBH growth history from the
observationally measured MBH spin and mass distribution, we first
generate a large number of mock MBHs for different settings of the
model parameters $(b,\ f_\c,\ \gamma)$. We assume that the final
masses of those MBHs follow the mass distribution of local AGNs
\citep{sch10} since almost all MBHs with spin measurements are at
redshift $z < 0.3$.\footnote{Note that currently no estimate is
available for the distribution of the final masses of those active
MBHs. We assume the mass function of local AGNs is close to the final
MBH mass distribution.  However, if the sample is sufficiently large,
one may simultaneously obtain the final MBH mass distribution and
constrain the MBH growth.} 

We randomly generate $4000$ MBHs over the logarithmic final mass
ranging from $10^6$ to $10^{10}M_\odot$. Then we adopt the  accretion
model described in Section~\ref{sec:spinevol} to calculate the spin
evolution curves for these $4000$ MBHs with each given set of
parameters ($f_\c$, $b$, $\gamma$), and take these spin evolution
curves as templates. Using the mass distribution function as the
weight, we randomly select mock AGNs and thus the AGN properties from
those templates, including the MBH mass and spin at the `observation'
time.

For the spin curve calculation, we set the initial masses of those
MBHs randomly distributed from $10^4$ to $10^5 M_\odot$ in the
logarithmic space and the initial spins are randomly distributed from
$0$ to $1$.\footnote{ Setting all the initial spins to $0$ or $1$ does
not introduce any significant difference as those initial information
are quickly washed out with the growth of MBHs.} In the
coherent-accretion phase, we set the accretion rate as $\dot{m} =
0.3$. A different choice of the accretion rate at the
coherent-accretion stage makes little difference to the spin evolution
as suggested by Figure~\ref{fig:f1} and described in
Section~\ref{sec:spinevol}. We will further discuss in
Section~\ref{sec:discussion}, however, a different choice of $\dot{m}$
does cause some difference on the frequency of those MBHs accreting
via thin disc and thus emitting Fe K$\alpha$ line in the inner disc
region at the early stage of MBH growth. In the chaotic thin disc
accretion phase, we assume that the Eddington ratio $f_\edd$ in each
episode is a constant, and $\log f_\edd$ is randomly drawn from a
Gaussian distribution with mean $-0.8$ and standard deviation $0.5$.
We have checked and found that it makes little difference to our
results if $f_\edd$ is drawn from the distribution of local active
MBHs \citep{sch10}, but the time needed to select mock samples with
appropriate luminosity is much longer.

With these settings, we obtain the spin evolution curves for those
MBHs by solving Eqs.~(\ref{eq:dmdt}) and (\ref{eq:djdt}) or
(\ref{eq:dadt}).  For each MBH, we record the mass, spin, and
bolometric luminosity ($f_\edd L_\edd$) every $4.5 \times10^4$ yr
($\sim 10^{-4} t_{\rm Edd}$). For the $i$-th observed AGN with mass
$M_{i, \obs}$, spin $a_{i, \obs}$, and luminosity $L_{i, \obs}$ listed
in Table~\ref{tbl-1}, we randomly select an evolution curve and a
random moment $t_i$ in the curve for the period via thin disc
accretion, and then obtain the mass $M_{i, \rm mock}$ and bolometric
luminosity $L_{i, \rm mock}$ of a mock AGN at that moment by
interpolation. The masses of active MBHs in Table~\ref{tbl-1} are
mostly determined through the empirical virial mass estimators, which
may deviate from the true masses with a scatter of $0.3-0.4$\,dex
\citep[e.g.,][see also \citealt{Vestergaard06}]{she08, she11}.  The
bolometric luminosity is usually derived from a combination of the
luminosity at a specific band with the corresponding bolometric
correction.  The bolometric corrections for a specific optical band
usually have a scatter of 0.1-0.2 dex \citep[e.g.,][]{hopkins07}. In
addition, there are other uncertainties, such as those induced by
absorption and host contamination, which are difficult to accurately
consider. We therefore set an empirical uncertainty of 0.3 dex, and
select a mock AGN as a correspondence to the $i$th object in
Table~\ref{tbl-1} if $\left|\log (M_{i, \rm mock}/ M_{i,
\obs})\right|<0.3$, and $\left|\log (L_{i, \rm mock}/ L_{i,
\obs})\right|<0.3$. The spin $a_{i, \rm mock}$ at that moment is
obtained by interpolation. If these inequalities are not satisfied,
then we select another curve and repeat the above processes. For each
observed source, we generate $1,000$ mock objects. Choosing a larger
number does not affect our final results.  After going through all the
27 observed sources, we obtain $27\times1000= 27,000$ mock objects.
The we obtain the spin-mass distribution output from each model
according to these mock objects.

\begin{figure*}
\centering
\includegraphics[width=7in]{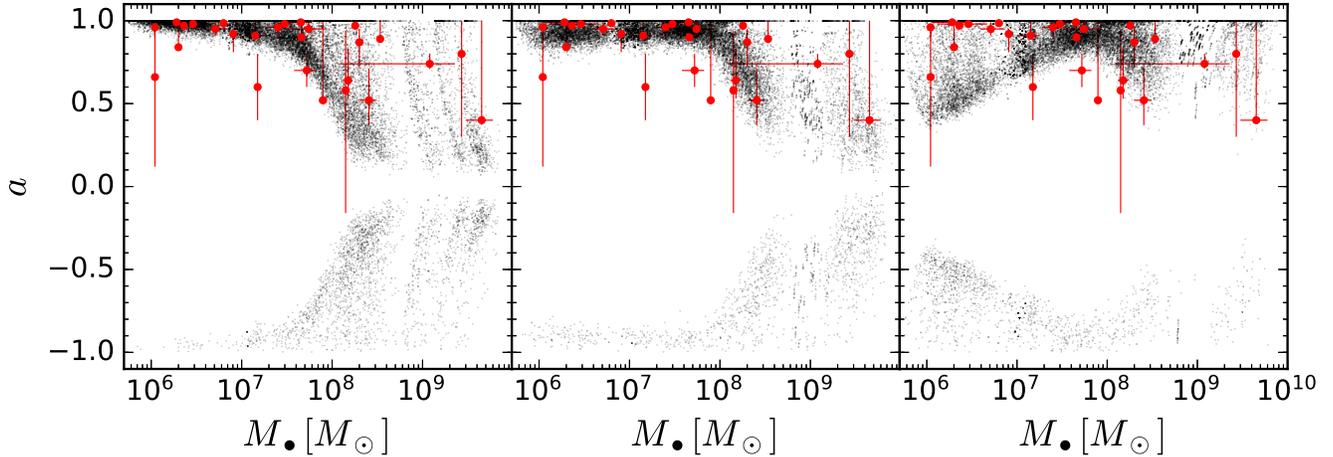}
\caption{Spins of the AGN mock samples (black dots) generated from the
two-phase accretion model and spins of the observed AGNs listed in
Table~\ref{tbl-1} (red filled circles with error bars). Left, middle,
and right panels represent $(f_\c,\ b,\ \gamma)=(0.48,\ 0.004,\
-0.5)$, $(0.46,\ 0.012,\ 0)$, and $(0.48,\ 0.014,\ 0.5)$,
respectively, corresponding to the cases marked by the star symbols in
Fig.~\ref{fig:f7}.
}
\label{fig-am}
\end{figure*}

%
For those objects with spin measurements listed in Table~\ref{tbl-1},
we assume a probability distribution function (PDF) for each source by
considering the spin measurement errors listed there (see similar
assumptions made in \citealt[][]{ses14}).  For objects given with
symmetric spin errors, a Gaussian PDF is assumed; for those given with
asymmetric errors, we assume the PDF is composed of two
half-Gaussians; for those given with lower limits,  the PDF is assumed
to have 90\% probability randomly distributed between $0.998$ and the
lower limit (90\% CL), and 10\% probability to lie between $-0.998$
and the lower limit. We randomly assign a spin to each of the $27$
object according to this assumed probability distribution for each
object, and obtain a spin-mass distribution for the observational
sample.  We compare this `observational' spin-mass distribution with
the model spin-mass distribution via the two-dimensional
Kolmogorov-Smirnov (2D-KS) test \citep{pre07}, and obtain the $P_{\rm
KS}$ value.  Although the 2D-KS test is not as rigorous as its
one-dimensional counterpart, it does return a p-value $P_\ks$ which
demonstrates the approximate probability that the two samples are
drawn from the same distribution. We repeat the above process for
$1000$ times and obtain $1000$ $P_{\rm KS}$ values for each model. We
take the median of those $P_{\rm KS}$ as the true $P_{\rm KS}$ value
for a given model. In such a way, we can investigate whether a model
with one set of $(f_\c, b, \gamma)$ matches the observational
spin-mass distribution better than another one by comparing the
$P_{\rm KS}$ values. 
%

\begin{figure}
\centering
\includegraphics[width=3.4in]{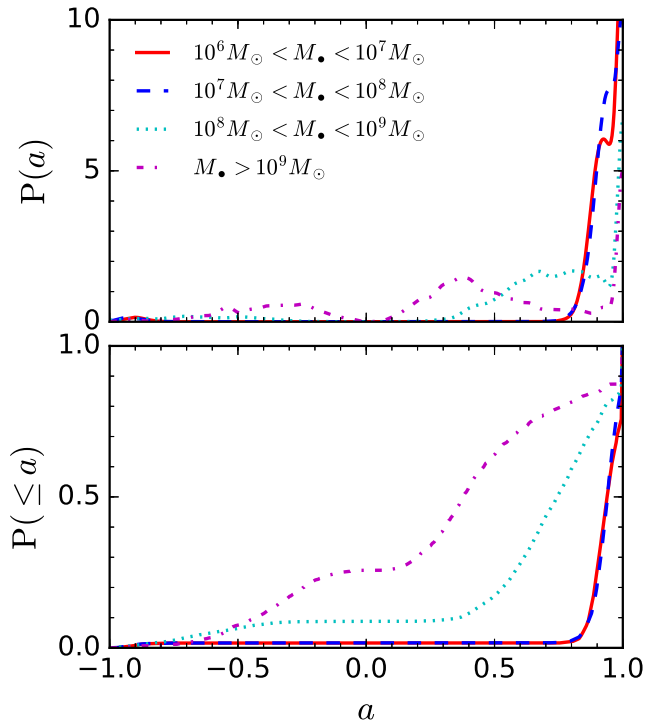}
\caption{Differential (top panel) and cumulative (bottom panel) spin
distribution of the mock AGN sample generated from the reference model
(black dots in the middle panel of Figure~\ref{fig-am}) in different
mass range, i.e., $10^6 M_\odot<M_\bh<10^7 M_\odot$ (red solid), $10^7
M_\odot<M_\bh<10^8 M_\odot$ (blue dashed), $10^8 M_\odot<M_\bh<10^9
M_\odot$ (cyan dotted), and $M_\bh>10^9 M_\odot$ (magenta dot-dashed).
}
\label{fig-pa}
\end{figure}

\begin{figure*}
\centering
\includegraphics[width=7in]{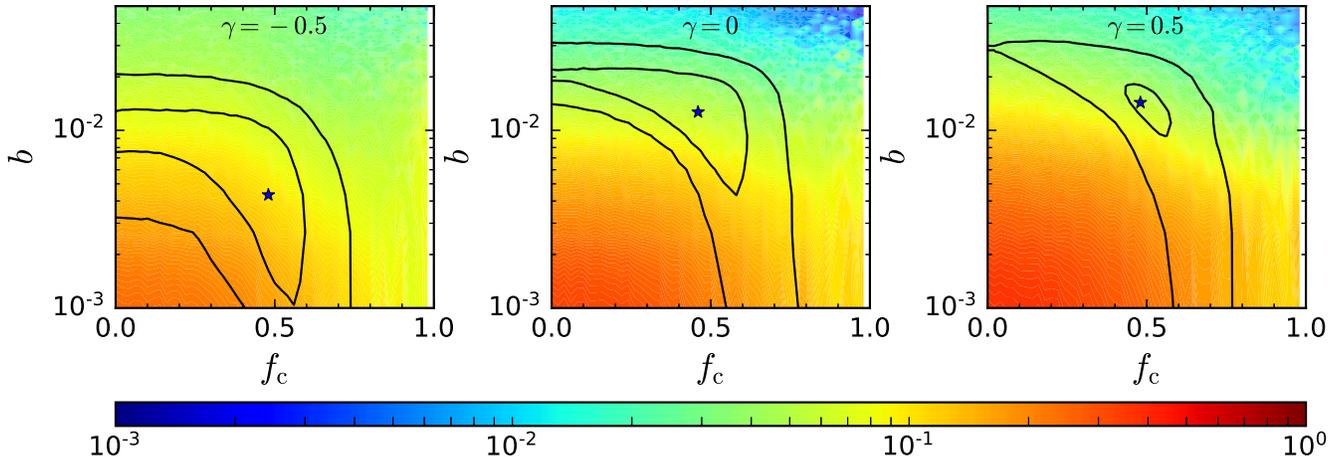}
\caption{Fraction of negative spins (i.e., retrograde accretion) in
the parameter space of $(f_\c,\ b)$ assuming $\gamma=-0.5$ (left
panel), $0$ (middle panel), and $0.5$ (right panel). The star symbols
mark the location of the model parameters that leads to the largest
$P_\ks$ value, and the black lines show $P_\ks=0.1$ and $0.01$, as
shown in Fig.~\ref{fig:f7}. The small white regions are for the
fraction of negative spins $<0.001$.
}
\label{fig-fneg}
\end{figure*}

\begin{figure}
\centering
\includegraphics[width=3.4in]{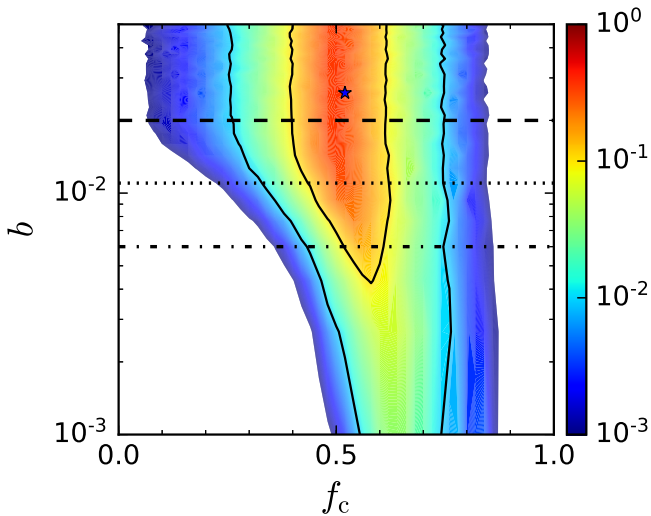}
\caption{Legends similar to that for the middle panel of
Fig.~\ref{fig:f7}, but considering the self-gravity of the thin disc
at the chaotic-accretion phase. Similarly, the star symbol marks the
location of the highest $P_\ks$ value $(f_\c, b)=(0.52, 0.026)$. The
black solid lines show $P_\ks=0.1$ and $0.01$, respectively. The
dashed/dotted/dot-dashed line marks the upper limit of $b$ for an MBH
with mass $M_\bh>10^6 M_\odot$/$10^7 M_\odot$/ $10^8 M_\odot$, above
which the disc is fragmented due to its self-gravity.  As seen from
this Figure, self-gravity of the disc may play a dominant role when
$b$ is substantially larger than one percent for $M_\bh>10^6 - 10^7
M_\odot$. }
\label{fig-msg}
\end{figure} 

\begin{figure}
\centering
\includegraphics[width=3.4in]{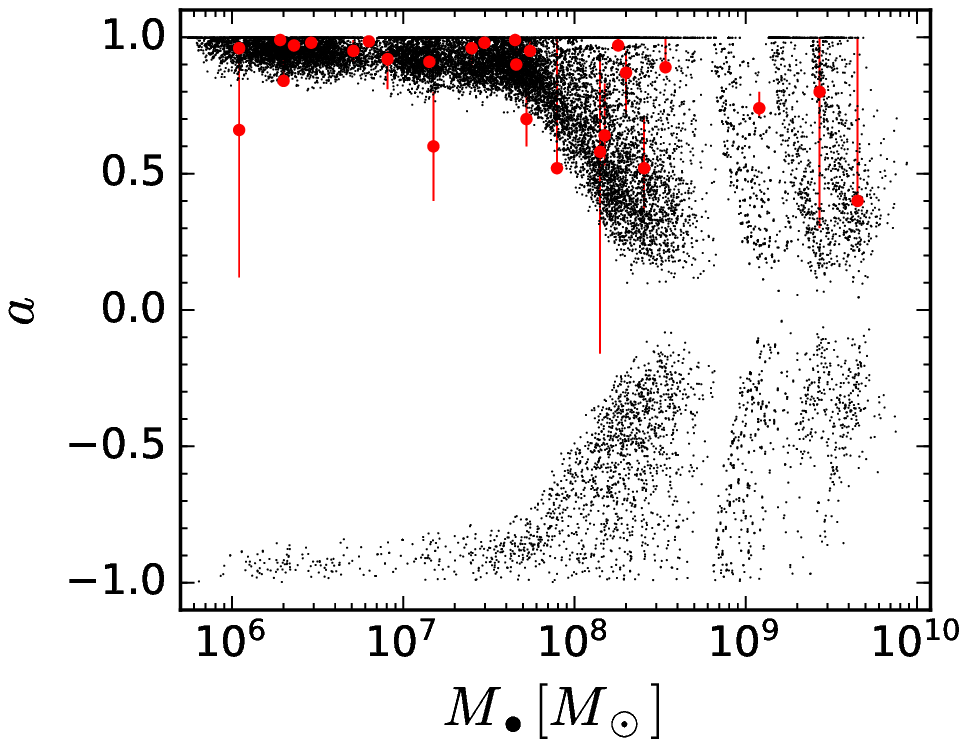}
\caption{Legend similar to Fig.~\ref{fig-am} but with consideration of
the self-gravity of discs at the chaotic-accretion phase.  }
\label{fig-amdata_msg}
\end{figure} 

We first consider the general case that the disc mass in each episode
of chaotic accretion $\propto M^{1+\gamma}_\bh$ [Eq.~(\ref{eq:mcl})].
Figure~\ref{fig:f7} shows the distribution of $P_\ks$ on the plane of
$b$ versus $f_\c$. For simplification, we set three different values
for $\gamma$, i.e., $-0.5$, $0$, and $0.5$ (from left to right
panels), to check whether the model with a positive or negative
$\gamma$ can match the observed spin distribution better, comparing to
the model with $\gamma=0$.  According to those models with different
$\gamma$ values as shown in Figure~\ref{fig:f7}, it can be excluded
with at least $99.9\%$ confidence level that the spin distributions of
the mock samples produced from those models with $f_\c>0.9$ and that
of the observed sample come from the same distribution,
which means that chaotic accretion is required and contributes at
least $10\%$ to the MBH final mass. Figure~\ref{fig:f7} also shows a
trend that a smaller $b$ allows for a larger $f_\c$. This is because a
smaller $b$ means a smaller disc and a more efficient decrease in spin
in the chaotic-accretion phase and a larger $f_\c$ implies less time
left for chaotic accretion. To reproduce the observed fraction of
intermediate spins (e.g., $\sim 0.4-0.8$), $b>0.05$ can also be
excluded at $99.9\%$ confidence level as it results in too many MBHs
with spins close to $1$.  Note that $f_\c\la 0.01$ cannot be excluded
(especially for $\gamma=0$; middle panel), because chaotic accretion
alone can also lead to the observed intermediate-to-high spin
distribution if the the disc mass (represented by $b$) is appropriate,
i.e., neither too large nor too small. As seen from
Figure~\ref{fig:f7} (and also Fig.~\ref{fig-am}), it appears that the
model with $\gamma=0$ matches the observations  ($P_\ks \sim 0.4$)
better than that with $\gamma= 0.5$ or $-0.5$. There also seems to be
a trend that a model with smaller $\gamma$ requires a smaller $b$ to
generate mock samples matching the observation. The reason is that a
smaller $\gamma$ should be coupled with a smaller $b$ (and vice versa)
in order to maintain an appropriate disc mass [$\propto b M_{\bullet}
\left(M_{\bh}/10^8M_\odot\right)^\gamma$] that can lead to the scatter
of spins in the mass range from a few times $10^7M_\odot$ to a few
times $10^8M_\odot$.

Figure~\ref{fig-am} shows the corresponding spin-mass distributions of
the mock samples given by those models with the largest $P_\ks$ value,
i.e., $(f_\c,\ b,\ \gamma)=(0.48,\ 0.004,\ -0.5)$ (left panel),
$(0.46,\ 0.012,\ 0)$ (middle panel; hereafter the reference model),
and $(0.48,\ 0.014,\ 0.5)$ (right panel), respectively. As seen from
this figure,  the distributions of the mock AGNs obtained from the
model with the highest $P_\ks$ value but different $\gamma$ show
different patterns, especially at the low mass end. For the model with
$\gamma=0.5$, the mock AGNs with $M_\bh \la 10^7 M_\odot$ have
intermediate to high spins $(\sim 0.5-1)$, while they mostly have high
spin $(\ga 0.8$) for the model with $\gamma=-0.5$.

\begin{figure}
\centering
\includegraphics[width=3.4in]{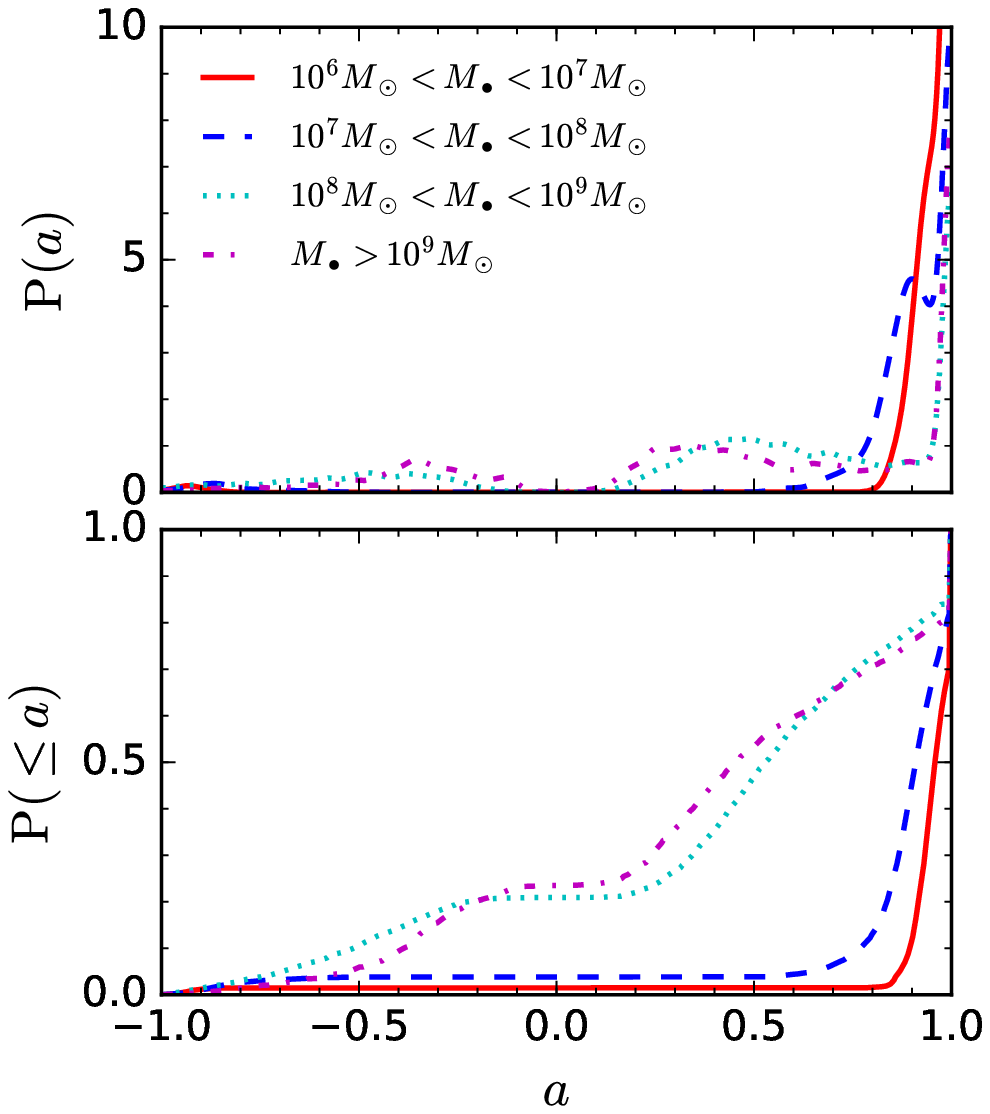}
\caption{Legend similar to Fig.~\ref{fig-pa} but considering the
self-gravity of discs at the chaotic-accretion phase.
}
\label{fig-pa_msg}
\end{figure} 

Figure~\ref{fig-pa} shows the differential and cumulative spin
distributions of those mock objects in different mass bins, in order
to illustrate how the spin distribution depends on the MBH mass. As
seen from this figure, higher mass MBHs ($>10^8M_\odot$) have
relatively low spins, which is a natural result of the two-phase
accretion model as chaotic-accretion phase normally leads to a fast
spin-down of heavy MBHs at their later growth stage (see
Figs.~\ref{fig-mbh}, \ref{fig-sg}, and \ref{fig:f6}).  It is also
prominent that a fraction of MBHs have negative spins. The reason is
that $J_\disc/J_\bh$ decreases with increasing MBH mass for a given
$b$ at the chaotic accretion phase [Eq.~(\ref{eq:jdojh})], and the
criterion for anti-alignment $\cos \beta <-J_\disc/2J_\bh$
\citep{kin05} can be satisfied in some cases.

Figure~\ref{fig-fneg} shows the fraction of mock AGNs that have
negative spins on the plane of $f_\c$ versus $b$,  obtained from those
models with $\gamma=-0.5$ (left panel), $0$ (middle panel), and $0.5$
(right panel), respectively. A model with smaller $b$ results in a
larger fraction of AGNs with negative spins, because $J_\disc/J_\bh$
is smaller, and the criterion for anti-alignment \citep[see][]{kin05}
can be more frequently satisfied. A larger fraction of negative spins
can also result from models with smaller $f_\c$ but the same $b$, as
the duration of the chaotic-accretion phase is longer and thus there
are more chances to have anti-aligned accretion disc. The model with
the largest $P_\ks$ value, i.e., $(f_\c,\ b)=(0.46,\ 0.012)$ results
in $\sim 5.5\%$ of the mock AGNs that have negative spins, i.e.,
roughly $1.5$ in $27$ objects. According to Figure~\ref{fig-pa},
however, one may note that the fraction of counter-rotating MBHs
depends on the MBH mass, i.e., the higher the MBH mass, the higher the
fraction (see Discussion part).

If we consider self-gravitated discs in the chaotic-accretion phase,
i.e., $M_\disc=\min(b M_\bh, M_\sg)$ if $\gamma=0$, then we can also
obtain the distribution of $P_\ks$ value on the plane of $f_\c$ versus
$b$ according to the same processes described above (as shown in
Figure~\ref{fig-msg}). Here $M_\sg$ is the maximum mass of an
accretion disc by considering self-gravity and disc fragmentation and
it is given in Appendix (Eq.~\ref{eq:msg}).  Similar to the previous
results, $f_\c$ is also constrained to be less than $\sim 0.85$ at
$\gtrsim 99.9\%$ confidence level, and the location of the model with
the largest $P_\ks$ value is $(f_\c,\ b)=(0.52,\ 0.026)$, roughly
consistent with the previous results. It seems that a large $b$ is now
allowed, apparently different from those models without consideration
of disc self-gravity. The reason for this difference is
straightforward, i.e., the disc mass is determined by $M_\sg$ for
those models with large $b$,  in which $b$ does not play a role
because $M_{\rm sg}  < bM_{\bullet} $. By setting $M_\sg=b M_\bh$, we
may obtain a critical  MBH mass below which $M_\disc = b M_\bh$ and
above which $M_\disc=M_\sg$. This critical mass is determined by $b$.
If we set the critical mass as $10^6 M_\odot$, it means that the
masses of discs around all MBHs with $M_\bh>10^6 M_\odot$ are limited
to $M_\sg$, and the corresponding $b$ value must be $\le 0.02$ (as
indicated by the dashed line in Fig.~\ref{fig-msg}). For a critical
MBH mass of $10^7 M_\odot$ or $10^8 M_\odot$, it gives an upper limit
of $b=0.011$ (dotted line) or 0.006 (dot-dashed line in
Fig.~\ref{fig-msg}).

Figure~\ref{fig-amdata_msg} shows the mock sample obtained from the
two-phase accretion model with the largest $P_\ks$ value by
considering self-gravity of the disc in the chaotic-accretion phase,
i.e., $(f_\c,\ b,\ \gamma) = (0.52,\ 0.026,\ 0)$, as a comparison to
the spin-mass distribution shown in the middle panel of
Figure~\ref{fig-am}. Figure~\ref{fig-pa_msg} shows the spin
distribution of the mock AGNs in Figure~\ref{fig-amdata_msg} for
different mass ranges. By comparing these two figures with
Figures~\ref{fig-am} and \ref{fig-pa}, we find that qualitatively the
results obtained with and without consideration of disc self-gravity
do not differ much.

\section{Discussion}
\label{sec:discussion}

\subsection{Different choices of the accretion rate in the coherent-
and chaotic-accretion phases}

\begin{figure}
\centering
\includegraphics[width=3.4in]{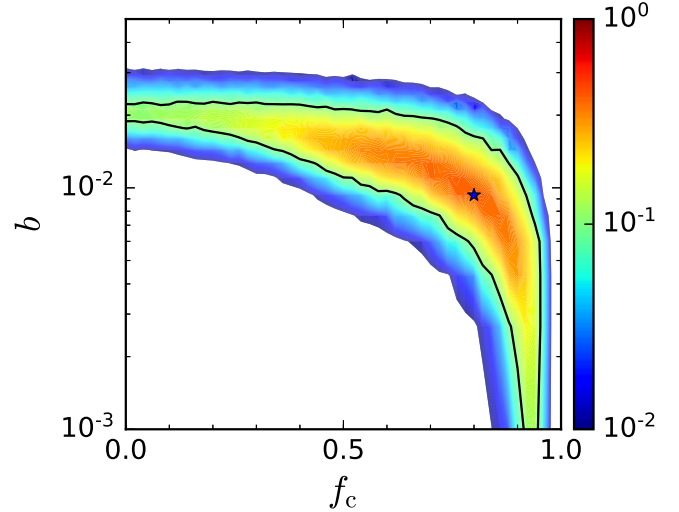}
\caption{Constraints on the parameters $b$ and $f_\c$ in the two-phase
accretion model from the available spin measurements of AGNs listed in
Table~\ref{tbl-1}. Here the accretion rate at the coherent-accretion
phase is set to be super-Eddington with $\dot{m}=100$. The star symbol
marks the location with the largest $P_\ks$ value $(f_\c,\ b)=(0.80,\
0.009)$.  The black contour shows $P_\ks=0.1$. }
\label{fig-md100}
\end{figure}

In our calculations presented in the previous section, the accretion
rate in the coherent-accretion phase is set to be a constant, i.e.,
$\dot{m}=0.3$. However, the accretion in the coherent phase can also
be super-Eddington. Therefore, we further check the case with
$\dot{m}=100$ and do similar calculations. Figure~\ref{fig-md100}
shows the distribution of $P_\ks$ obtained from models with $\gamma=0$
and $\dot{m}=100$ in the coherent-accretion phase. We find that those
models with large $f_\c$ (e.g., $0.9<f_\c <0.95$) here can still be
compatible with the observations, and the model with the largest
$P_\ks$ value is $(f_\c,\ b)= (0.80,\ 0.009)$. For comparison, models
with $f_\c>0.9$ are ruled out with a high confidence $99.9\%$ if
$\dot{m} = 0.3$  (middle panel of Figure~\ref{fig:f7}). The reason is
that the time for the coherent super-Eddington accretion phase is
short, and even if $f_\c=0.9$, for example, there is still 10\% of
chaotic thin disc accretion for the MBH growth, from which mock
objects can be selected to match the observational spin distribution.
In contrast, if $\dot{m}=0.3$ and $f_\c=0.9$, then the probability is
high to select mock objects in the coherent phase, when most MBHs are
maximally spinning, and thus this model overproduces MBHs with spins
close to $1$, especially when MBHs are large ($> 10^8 M_\odot$).

\begin{figure}
\centering
\includegraphics[width=3.4in]{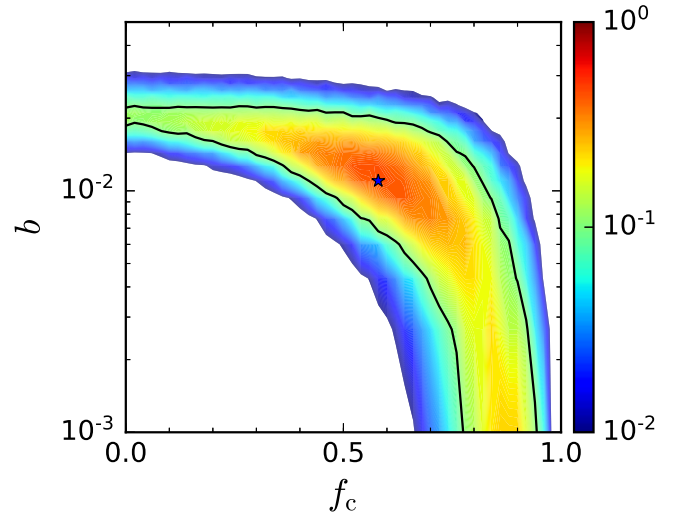}
\caption{Legend similar to Fig.~\ref{fig-md100}. However, here the
accretion rate in the coherent-accretion phase is set to be the
possible upper limit for the accretion rate given by \citet{lev18}.
The star symbol marks the location of the largest $P_\ks$, i.e.,
$(f_\c,\ b)=(0.58,\ 0.011)$. }
\label{fig-mdconst}
\end{figure}

In addition, the accretion rate in the coherent stage may change with
time.  \citet{lev18} have shown that the accretion rate of MBHs at
early epochs can not exceed 
\be
\dot{M}_{\rm acc} \sim 20 \left(\frac{\sigma}{350\ {\rm
km\,s^{-1}}}\right)^4 M_\odot \ \rm{yr^{-1}},
\ee
where $\sigma$ is the stellar velocity dispersion. Considering the
$M_\bh-\sigma$ relation, i.e., $M_{\bh, f}$ roughly proportional to
$\sigma^{4}$ \citep[see][]{tre02}, we have 
\be
\dot{M}_{\rm acc} \sim 1.4 \frac{M_{\bullet,f}}{10^8M_\odot}
M_\odot{\rm \,yr}^{-1} \sim 0.65 \frac{M_{\bullet,f}}{M_{\bullet}}
\dot{M}_\edd(M_\bh). \nonumber \\
\ee
The accretion rate at the early stage ($M_\bullet \ll M_{\bullet,f}$)
is significantly higher than the Eddington rate and scales with the
MBH final mass. We therefore use the above $\dot{M}_{\rm acc}$ as the
accretion rate in the coherent phase to perform similar calculations
as in Section~\ref{sec:constraint}. Figure~\ref{fig-mdconst} shows the
distribution of $P_\ks$ on the plane of $f_\c$ versus $b$. The
location for the largest $P_\ks$ is $(f_\c,\ b)= (0.58,\ 0.011)$,
which is in between that obtained by assuming $\dot{m}=0.3$ (middle
panel of Fig.~\ref{fig:f7}) and $\dot{m}=100$ (Fig.~\ref{fig-md100}),
and only slightly differs from them. This is a natural result because
a constant $\dot{M}_{\rm acc}$ in the coherent phase for an MBH with
given $M_{\bullet,\rm f}$ means a decreasing $\dot{m}$ with increasing
MBH mass, from super-Eddington (e.g., $\dot{m}=100$ at $M_\bh
<10^6M_\odot$) to sub-Eddington (e.g., $\dot{m} \sim 0.3$ when $M_\bh$
is close to $M_{\bullet,\rm f}$).

In our models, the Eddington ratio is also assumed to be constant
within each accretion episode and randomly generated for each episode
in the chaotic accretion phase.  It may be more realistic to assume a
time-varying Eddington ratio (or accretion rate) for each episode,
i.e., a power-law decay of $f_\edd$ or accretion rate
\citep[e.g.][]{YLK05, Hopkins06, Aversa15}. However, choosing a
different Eddington ratio makes little difference to the spin-mass
evolutionary curve, although the accretion time and the light curve
can be quite different. It has an equivalent effect by assuming a
constant $f_\edd$ in a single chaotic accretion episode but varying
$f_\edd$ in different episodes if the number of accretion episode is
large.
%

 It is apparent that $f_c>0.95$ can be excluded at $\sim 99\%$
confidence level although different settings on the accretion rates
for the coherent-accretion phase may result in different constraints
on model parameters, which means chaotic accretion is necessary.  Pure
chaotic accretion is still allowed, and in this case $b$ is all
constrained to be around 0.01, i.e., $0.007<b<0.03$.

To close this sub-section, we note here that the simple two-phase
accretion model adopted in the present paper has some limitations as
it ignores the possibility that the accretion history could be much
more complicated. For example, the accretion histories may be
different for those MBHs in different environments, activated at
different redshifts, or with different masses
\citep[e.g.,][]{2018MNRAS.477.3807F}. In a much more comprehensive
cosmological co-evolution model for MBHs (both masses and spins) and
galaxies, one may be able to consider all those complications and
obtain detailed constraints on the MBH accretion histories
\citep[e.g., see][]{ses14, Lapi06, Shen09}. With such a model, one can
also simultaneously obtain the QSO luminosity function, the clustering
of QSOs, and compare them with observations. However, there are also
many more uncertainties because of poor understanding of many physical
processes involved in the co-evolution of MBHs and galaxies.

\subsection{Mass-dependence of $f_\c$ and $b$}

One of the settings in our previous calculations is that $f_\c$ and
$b$ are both the same for all MBHs. This may not be the fact as $f_\c$
(and/or $b$) may depend on the MBH (final) mass. 

In order to test this possibility, on the one hand, we assume a simple
power-law form $f_\c=f_0 (M_{\bh,\rm f}/10^8 M_\odot )^x$, and fix $b$
at $0.012$ (the $b$ value of the reference model). By matching the
observational spin distribution through the 2D-KS test, we find that
the parameters of the model with the largest $P_\ks$ value are now
$(f_0,\ x)=(0.38,\ 0.06)$. If alternatively adopting $b=0.02$ or
$b=0.005$, we obtain $(f_0,\ x)=(0.10,\ 0.16)$ or $(0.46,\ 0)$. These
results imply that there is no necessity to assume an $f_\c$ dependent
on the final MBH mass to match the currently available spin
measurements.  However, if $f_\c$ for high-mass MBHs is much smaller
than that for low-mass ones (with $b$ fixed at the value of the
reference model), the fraction of slowly spinning MBHs at the high
mass end would be substantially larger, while if $f_\c$ for low-mass
MBHs is relatively small compared with that for high-mass ones, there
will be little change in the resulting spin distribution at the low
mass end provided that $b$ is not too small (e.g., $b > 10^{-3}$).

On the other hand, if $b$ is, not necessarily monotonically, dependent
on the final MBH mass, then the resulting MBH spin distribution may be
significantly different from the reference model. Assuming $b \propto
M^{\kappa}_{\bullet,f}$ is similar to the setting described by
Equation~(\ref{eq:mcl}), and therefore, its effects on the MBH spin
distribution can be seen from Figure~\ref{fig-am}.  However, if $b$ is
less than $0.01$ for MBHs with mass $\ga 10^8 M_\odot$ and $\la
10^6-10^7 M_\odot$ but is $\sim 0.01$ for MBHs with masses in between,
then the spins of MBHs at both the high- and low-mass ends can be
broadly distributed with a significant fraction locating at close to
$0$ if $f_\c\la 0.5$.  One example case of small $b$ for $10^6-10^8
M_\odot$ MBHs might be that the accretion of tidally disrupted stars
contributes significantly to the growth of low mass MBHs, as discussed
in a separate paper \citep[e.g.,][]{ZLZ18}, in which $b\rightarrow 0$
in the phase of accreting tidally disrupted stars, and will lead to
low-spin MBHs at the low mass end.

\subsection{Quantitative constraints and parameter degeneracies}

In this paper, we have applied the simple 2D-KS test to compare the
spin distribution of mock samples with the observational ones and
obtain constraints on the MBH accretion histories. According to the
results described above, apparently significant degeneracies exist
among the constraints on the model parameters. For example, the models
assuming super-Eddington accretion in the coherent-accretion phase can
also give good matches to the observations (see Figs.~\ref{fig-md100}
and \ref{fig-mdconst}), and it is not easy to distinguish them from
those models assuming thin disc accretion in the coherent-accretion
phase. In principle, one may apply the Bayesian technique to obtain
constraints on the MBH growth and possibly break some of the
degeneracies among the model parameters as that discussed in
\citet[][]{ses14}. However, many of the current spin measurements
listed in Table~\ref{tbl-1} only give lower limits and are not
sufficiently accurate, which prevents a concrete Bayesian analysis
without additional assumptions on the probability distribution of each
measured spin value.  There will be many more MBH spins that can be
measured accurately with future X-ray telescopes such as the Advanced
Telescope for High Energy Astrophysics (Athena), Hitomi, the Large
Observatory for X-ray Timing (LOFT), and the Enhanced X-ray Timing and
Polarimetry (eXTP). With such spin measurements, one may generate mock
observations according to the co-evolution model(s) for MBHs and
galaxies with more detailed parameterized accretion histories of MBHs,
and investigate MBH growth histories by using more concrete Bayesian
statistics. One could also do some simulations according to the
`observations' of those future X-ray telescopes to demonstrate whether
some of the parameter degeneracies can be broken and more rigorous
constraints on MBH growth can be obtained, which is deferred to a
future work.

\subsection{Radiative efficiency of MBHs}

\begin{figure}
\centering
\includegraphics[width=3.4in]{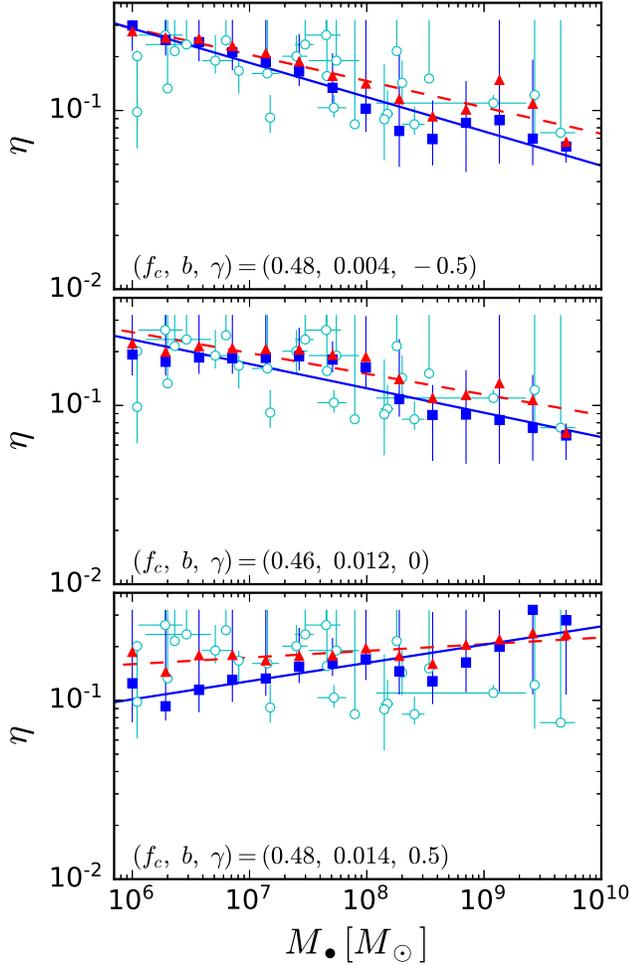}
\caption{Mean (red triangles) and median (blue squares) radiative
efficiencies for the mock AGNs in different MBH mass bins. The mock
AGN samples are obtained from those models with the largest $P_{\rm
KS}$ value for the cases of $\gamma = -0.5, 0$, and $0.5$, i.e.,
$(f_{\rm c},\ b,\ \gamma)= (0.48,\ 0.004,\ -0.5)$ (top panel),
$(0.46,\ 0.012,\ 0)$ (middle panel), and $(0.48,\ 0.014,\ 0.5)$
(bottom panel) (as shown by the black dots in the left, middle, and
right panels of Fig.~\ref{fig-am}), respectively. Cyan circles show
the radiative efficiency of individual sources with spin measurements
and they are directly converted from the spin measurements assuming
the standard thin disc accretion model. The blue solid (red dashed)
lines are the best linear fits to the blue squares (red triangles) in
each panel, and the blue error bars represent the 16$th$ and 84$th$
percentiles.
}
\label{fig-eta}
\end{figure}

\begin{figure}
\centering
\includegraphics[width=3.4in]{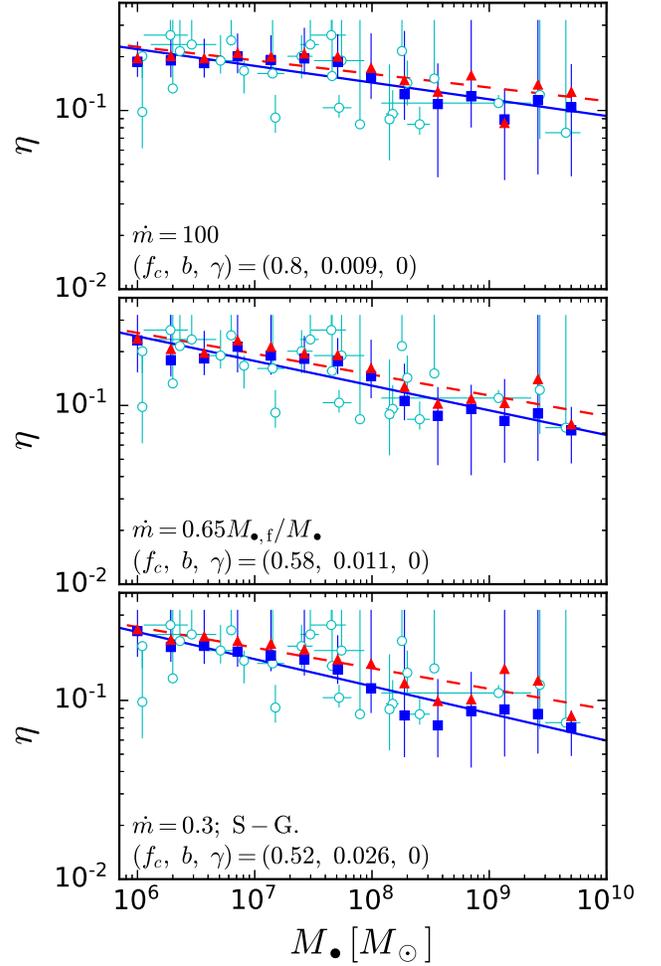}
\caption{Legends similar to Fig.~\ref{fig-eta}, but the mock AGN
samples are obtained from those models with the largest $P_{\rm KS}$
value for the case with $\dot{m} = 100$ (top panel), $\dot{M}_{\rm
acc}\propto \sigma^4$ (middle panel), and $\dot{m}=0.3$ with
consideration of the self-gravity of discs in the chaotic-accretion
phase.  
}
\label{fig-eta2}
\end{figure}

\begin{table*}
\begin{center}
\caption{Best parameters to fit the binned mean and median efficiencies with 
$\log\eta= \eta_8 + \zeta \log(M_\bh/10^8 M_\odot)$. }
\label{tbl-2}
\renewcommand{\arraystretch}{1.5}
\begin{tabular}{cccccc}
\tableline
\tableline
\multirow{2}{*}{$\eta_{8,\rm mean}$} & \multirow{2}{*}{$\zeta_{\rm mean}$} &
\multirow{2}{*}{$\eta_{8,\rm med}$} & \multirow{2}{*}{$\zeta_{\rm med}$}   & \multicolumn{2}{c}{model parameters}  \\ \cline{5-6}
& & & & coherent phase ($\dot{m}$) & chaotic phase $(f_\c, b, \gamma)$ \\
\tableline
$-0.84\pm0.14$ 	& $-0.15\pm0.02$ 	& $-0.93\pm0.13$ 	& $-0.20\pm0.02$  	& $0.3$ & $(0.48, 0.004, -0.5)$ \\
$-0.82\pm0.13$ 	& $-0.12\pm0.02$ 	& $-0.90\pm0.14$ 	& $-0.14\pm0.02$  	& $0.3$ & $(0.46, 0.012, 0)$ \\
$-0.72\pm0.08$ 	& $0.04\pm0.01$ 	& $-0.79\pm0.15$ 	& $0.10\pm0.02$  	& $0.3$ & $(0.48, 0.014, 0.5)$ \\
$-0.82\pm0.13$ 	& $-0.12\pm0.02$ 	& $-0.92\pm0.12$ 	& $-0.15\pm0.01$  	& $0.3$  & $(0.52, 0.026, 0),$ {\rm self-gravitating} \\
$-0.80\pm0.14$ 	& $-0.08\pm0.02$ 	& $-0.84\pm0.11$ 	& $-0.09\pm0.01$  	& $100 $& $(0.80, 0.009, 0)$    \\
$-0.83\pm0.13$ 	& $-0.12\pm0.02$ 	& $-0.89\pm0.12$ 	& $-0.14\pm0.01$  	& $0.65 \frac{M_{\bullet,\rm f}}{M_\bullet}$ & $(0.58, 0.011, 0)$    \\
\tableline
\end{tabular}
\end{center}
\tablecomments{Columns (1) and (2) are for the the mean efficiencies (red dashed lines in 
Figs.~\ref{fig-eta} and \ref{fig-eta2}), while columns (3) and (4) are for median ones 
(blue solid lines in Figs.~\ref{fig-eta} and \ref{fig-eta2}). 
}
\end{table*}

The mock samples (in Fig.~\ref{fig-am}) that match the observations
best are selected from the thin disc accretion stage, and thus the
radiative efficiency of each MBH in those samples can be directly
estimated from the MBH spin. We also calculate the mean and median
efficiencies in each mass bin according to the mock samples obtained
for the cases with $\gamma=-0.5$, $0$, and  $0.5$, respectively (see
Fig.~\ref{fig-eta}). We use a simple power-law model to fit the
possible relation, if any, between efficiency and MBH mass, i.e., 
\be
\log\eta = \eta_8+\zeta \log(M_\bh/10^8 M_\odot),
\ee
and the best-fit parameters $\eta_8$ and $\zeta $ are listed in
Table~\ref{tbl-2}.  For all those three $\gamma$ cases, it appears
that a weak correlation exists between the mean (or median) efficiency
and the MBH mass, i.e., $\eta \propto M^\zeta_\bh$ with $|\zeta|<0.2$.
For other models, i.e., with $\dot{m}=100$ or $\dot{M}_{\rm acc}
\propto M_{\bullet, \rm f}$ at the coherent-accretion phase, or with
self-gravitated disc considered, we plot the efficiency-mass relation
in Figure~\ref{fig-eta2} and list the best-fit parameters to this
relation in Table~\ref{tbl-2}.

Almost all those models result in a weak anti-correlation between
efficiency and mass, except that the model with $\gamma=0.5$ results
in almost no correlation. The result that the efficiency declines with
the MBH mass  is in contradiction with the positive correlation found
by \citet{DL11}, i.e., $\eta \propto M_\bullet^{0.5}$.  However, we
note here that the correlation between $ \eta $ and $M_\bullet$ found
in \citet{DL11} can be due to selection biases as pointed out in
\citet[][]{Wu13} (see also \citealt{Raimundo12}). It is also possible
that the sample of the current spin measurements is heterogeneous and
biased from the parent sample as further detailed in
Section~\ref{sec:agnsample}.

The mean efficiencies above are estimated from the mock samples by
averaging over a number of mock MBHs within each mass bin. 
We may also estimate the mean accreted-mass-weighted efficiency
($\left< \eta \right>_{\rm m}$) for the whole accretion process of all
mock MBHs, by considering the number density of MBHs with different
masses. The mean efficiency is also mostly contributed by MBHs with
mass around $10^8M_\odot$ as they dominate the MBH mass density.
For the model with $(f_\c,\ b,\ \gamma)=(0.46,\ 0.012,\ 0)$, we obtain
$\left< \eta \right>_{\rm m}= 0.13$. For those models with $(f_\c,\
b,\ \gamma) =(0.48,\ 0.004,\ -0.5)$, $(0.48,\ 0.014,\ 0.5)$, the model
considering disc self-gravity with $(f_\c,\ b,\ \gamma)= (0.52,\
0.026,\ 0)$, the model considering super-Eddington accretion
($\dot{m}=100$) in the coherent accretion phase $(f_\c,\ b,\ \gamma)=
(0.80,\ 0.009,\ 0)$, and the model with an upper limit on the
accretion rate according to \citet{lev18} with $(f_\c,\ b,\ \gamma)=
(0.58,\ 0.011,\ 0)$, we obtain $\left< \eta \right>_{\rm m}= 0.15$,
$0.09$, $0.12$, $0.05$, and $0.11$, respectively.
For most of those models, the resulting value of $\left< \eta
\right>_{\rm m}$ is consistent with the global constraint ($\sim
0.09-0.12$) obtained by comparing the local MBH mass density with the
accreted MBH mass density via AGNs and QSOs in a number of references
(e.g., \citealt[][and references therein]{yu02, Elvis02, Marconi04,
Shankar04, Shankar09, Raimundo09, Raimundo12, zha12, ZL17}).  One may
also note that a few recent works obtained a smaller mean efficiency
of $\sim 0.05-0.06$ \citep[e.g.,][] {shankar13, ueda14}, which might
be caused by adopting a higher local MBH mass density. However, if one
considers the sample bias when estimating the local MBH density via
different scaling relations, i.e., $M_\bullet-\sigma$ and $M_\bullet-
L$ relations \citep[e.g.,][]{bernardi07, shankar16}, the value $\sim
0.05-0.06$ might be an underestimate and not necessarily contradict
with the mean efficiency obtained from those models without assuming
super-Eddington accretion.  The mean value of the radiative efficiency
($\sim 0.05$) resulting from the model setting a super-Eddington
accretion rate of $\dot{m}=100$ in the coherent-accretion phase,
although consistent with the estimates by \citet{shankar13} and
\citet{ueda14}, seems lower than the global constraints obtained by
most authors. With more measurements on MBH spins in the future, it
would be possible to constrain whether most MBHs experienced a
significant super-Eddington accretion by combining the estimate of
mean efficiency through an independent method.

\subsection{Possible cosmic evolution of MBH spins?}
\label{subsec:spincosmevol}

The two-phase model constrained above suggests that low mass MBHs spin
faster than high mass ones as a combined effect of two factors: 1) the
de-spin due to the late stage chaotic accretion, and 2) the more
efficient de-spin of those MBHs with $M_\bullet \ga 10^8M_\odot$ in
the chaotic accretion phase. Although the cosmic evolution of MBHs is
not considered in the above calculations, the two-phase model may
imply a cosmic spin evolution of the active MBH population. A more
comprehensive model should include both the MBH spin and mass
evolution in the co-evolution model for MBHs and galaxies
\citep[see][]{vol05, ber08, Dubois14_1, Dubois14_2, ses14}, though the details of the fueling
to MBHs are still not well understood. With upcoming spin measurements
and better determined spin distribution, it is possible to combine the
constraints on the assembly history of MBHs with the co-evolution
model for MBHs and galaxies to investigate the cosmic evolution of the
spins of MBHs as a population. In such a cosmological evolution model
of MBHs, the clustering of active MBHs can be used to put strong
constraints on the lifetime and evolution of accretion histories
\citep[e.g.,][]{Aversa15}.  For example, if the accretion rate is too
high, with an extremely large $\dot{m}$, and the lifetime of QSOs is
too short, the inferred clustering would be too low to be consistent
with the observational one; if the lifetime is too long, the inferred
clustering would be too high to be consistent with the observation.
Therefore, the clustering of active MBHs can also put further
constraint on MBH spins because larger spins mean longer lifetime in
order to fit the observationally determined QSO luminosity functions.

\subsection{AGN sample with spin measurements}
\label{sec:agnsample}

Those models with largest $P_\ks$ value all result in a non-negligible
fraction of MBHs with negative spins. In the reference model [$(f_\c,
b, \gamma)=(0.46, 0.012, 0)$; shown in the middle panel of
Fig.~\ref{fig-am}], for example, about $5.5\%$ of the mock objects
are accreting via discs counter-rotating around their central MBHs.
However, all the active MBHs in Table~\ref{tbl-1} have positive spins.
One reason might be the small sample size of MBHs with spin
measurements. In Table~\ref{tbl-1}, $18$ objects have masses in the
range from $10^6$ to $10^8 M_\odot$, and the other $9$ have masses $>
10^8 M_\odot$.  The fraction of negative spins resulting from the
reference model is $1.5\%$ in the mass range $10^6-10^8 M_\odot$,
while it increases to $\sim 13.3\%$ for $> 10^8 M_\odot$ MBHs.
According to the Poisson statistics, the probability to observe $k$
counter-rotating MBHs is given by
$P(k)=\frac{e^{-\lambda} \lambda^k}{k!}$,
if the expected number of counter-rotating MBHs is $\lambda$, where
$k=0, 1, 2, ...$. Then the probability of non-detection of negative
spins among the $18$ objects with mass of $10^6-10^8 M_\odot$ is quite
high, i.e., $P(0) \simeq 0.76$, and the probability not to detect
negative spins among the $9$ objects with mass $>10^8 M_\odot$ is
$P(0) \simeq 0.30$ (see Fig.~\ref{fig-pa}).  Therefore, non-detection
of a negative spin  appears not a serious problem as the size of the
currently available spin sample is small. One may note that a recent
spin measurement of 1H 1934-063, a narrow-line Seyfert 1 galaxy, gives
$a<0.1$ \citep{2018arXiv180206056F}, which seems to be consistent with
a non-spinning or even a counter-rotating MBH. However, this
measurement only gives an upper limit and is not included in the above
2D-KS tests in which cumulative spin distributions (with spin smaller
than a value of $a$) were considered.  

The other reason might be that the sample listed in Table~\ref{tbl-1}
is incomplete.  Though it is still not clear whether the sample is
biased or not, some authors \citep[e.g.,][]{bre13} indeed pointed out
that it is more likely to detect high spin MBHs by applying the X-ray
reflection spectroscopy method.  The inner disc radius of a high spin
and/or prograde system is smaller than that of a low spin and/or
retrograde system, and thus relativistic effects are more prominent.
High signal-to-noise ratio spectroscopy is also needed in order to
measure the spin. This means the targeted object should be bright
enough, which may be easier satisfied for those MBHs radiating
efficiently with high spins. 

The limited sample size and the possible bias present caveats in
constraining the MBH growth history described above. Future  X-ray
telescopes, such as  Athena, Hitomi, LOFT, and eXTP, may accurately
measure several hundred or more MBH spins without or with less bias,
and thus may provide stronger constraints on the growth history of
MBHs. 

There are only three MBHs in Table~\ref{tbl-1} having mass $M_\bh>10^9
M_\odot$, i.e., H 1821+643 with $a>0.4$, Q 2237+305 with $a=
0.74^{+0.06} _{-0.03}$, and SDSS J094533.99+100950.1 with $a=
0.8^{+0.2}_{-0.5}$. It is important to check whether the constraints
obtained above are significantly affected by these three sources. We
therefore exclude these three sources and repeat the calculation, and
we find that the obtained constraints do not differ much from the
above results. It is worth noting that two important factors lead to
the constraints on the MBH accretion history. First, the spin
distribution for MBHs with mass ranging from $10^6-10^8 M_\odot$ in
Table~\ref{tbl-1} is narrow and the majority of the samples have spins
$\ga 0.8$, which suggest that the disc mass in the chaotic-accretion
phase, if any, cannot be too small. If it is too small (e.g., $b<
0.003$), then the spin distribution at this mass range cannot be that
narrow. We have checked that if only using those spin observations for
MBHs with mass $<10^8M_\odot$, we still obtain similar constraints on
$(f_\c, b)$, though the values of the largest $P_{\rm KS}$ does
decrease somehow and the constraints  become slightly less strong.
Second, about one third of the observed objects have masses $M_\bh \ga
10^8 M_\odot$ and spins broadly distributed, i.e. $a\sim 0.5-1$. The
existence of these objects indicates the amount of the MBH mass
coming from chaotic-accretion phase and suggests that chaotic
accretion phase is significant for the MBH growth and spin evolution.

\section{Conclusions}
\label{sec:summary}

In this paper, we study the spin evolution of MBHs under the
assumption that MBHs experienced two accretion phases, with an initial
phase of coherent-accretion via either the standard thin disc or
super-Eddington accretion, followed by a second phase of chaotic
accretion via the standard thin disc. As illustrated by many authors
in previous works, if the first phase dominates the MBH growth, then
most MBHs should be quickly spun up, to close to the maximum spin
value (e.g., $\sim 0.998$);  if the chaotic-accretion phase is
important, MBHs with mass $\ga 10^8 M_\odot$ may be significantly spun
down at their late growth stage. If the chaotic accretion phase
dominates the growth of MBHs and the coherent accretion is negligible,
we further find that the spins of those MBHs may quickly reach a
quasi-equilibrium state at their early growth stage and this state
ends up when the disc size becomes smaller than the warp radius of
non-equatorial disc(s). The value of the spin at the quasi-equilibrium
state is roughly determined by the mean ratio of the disc mass to the
MBH mass in the chaotic-accretion phase, i.e., the smaller this ratio,
the smaller the equilibrium spin value. Therefore, the spin
distribution of those MBHs with mass $\sim 10^6-10^8 M_\odot$ is
mainly determined by the mean (or distribution of) ratio of the disc
mass to the MBH mass in chaotic accretion episodes. 

Utilizing the spin evolutionary models studied in this paper, we
further investigate how the constraints on the MBH growth histories
can be obtained from the latest available spin measurements via the
X-ray reflection spectroscopy. We find that MBHs should experience a
chaotic-accretion phase with many accretion episodes, and on average
the mass accreted within each episode is roughly $1$-$2$ percent of
the MBH mass or less. The total amount of mass accreted in the chaotic
phase is at least $5-20$ percent of the final MBH mass. MBHs with
masses $\ga 10^8M_\odot$ appear to have intermediate-to-high spins
($\sim 0.5-1$), while MBHs with lower masses ($\sim 10^6 -10^8
M_\odot$) appear to have higher spins ($\ga 0.8$). On average, the
radiative efficiencies of those active MBHs appear  to slightly
decrease with increasing MBH masses. This indicates that the
correlation between the radiative efficiency and the MBH mass, if any,
is weak. The mean radiative efficiency of active MBHs is $\sim
0.09-0.15$, consistent with the global constraints by comparing the
accreted mass density with the local MBH mass density. 

\acknowledgements
We thank Qingjuan Yu for helpful discussions on and contributions to
various aspects presented in this paper. We thank the referee for
helpful comments and suggestions.  This work is partly supported by
the National Natural Science Foundation of China (Grant No. 11873056,
11690024, and 11390372), the National Key Program for Science and
Technology Research and Development (Grant No. 2016YFA0400704), and
the Strategic Priority Program of the Chinese Academy of Sciences
(Grant No. XDB 23040100).


\appendix \section{Inner boundary of the standard thin accretion disc
around an MBH} \label{sec:app1}

For an MBH accreting with $f_\edd \lesssim 1$,  the accretion disc is
assumed to be geometrically thin, optically thick, and described by
the standard thin disc model \citep[e.g.,][]{sha73, NT73}. Then the
inner boundary of the disc is \citep{bar72}
\be
R_{\rm in} = R_{\rm ISCO} = {3+Z_2 \mp[(3-Z_1)(3+Z_1+2Z_2)]^{1/2} },
\nonumber \\
\label{eq:rin}
\ee
with 
\be
Z_1 &=& 1+(1-a^2)^{1/3}[(1+a)^{1/3}+(1-a)^{1/3}] ,  \\
Z_2 &=& (3a^2+Z_1^2)^{1/2},
\ee
where $a=\left|a\right|$ when the disc is prograde rotating,
$a=-\left|a\right|$ when the disc is retrograde rotating, the upper
(lower) case of `$\mp$' (or `$\pm$') sign represents the prograde
(retrograde) orbit, and the same afterwards.

The general form of the specific energy $E$ and angular momentum
$\Phi$ as a function of radius $r$ (in unit of $r_{\rm g} =
GM_\bullet/c^2$) of a circular orbit are given by \citep{bar72}
\be
E(r)=\frac{r^{3/2}-2r^{1/2} + a}{r^{3/4}(r^{3/2}-3r^{1/2} +
2a)^{1/2}}, 
\label{eq:ein}
\ee
\be
\Phi(r) = \pm \frac{(r^2 - 2ar^{1/2}+a^2)}{r^{3/4}(r^{3/2}-3r^{1/2} +
2a)^{1/2}}.
\label{eq:phiin}
\ee
Then the specific energy and angular momentum at $R_{\rm ISCO}$ can be
obtained by setting $r=R_{\rm ISCO}$.

\section{inner disc boundary of thick discs} \label{sec:app2}

For super-Eddington accretion, we apply the logarithmic dependence of
Eddington ratio on the accretion rate given by \citet{min00}, i.e.,
\be 
f_\edd=\left\{ 
\begin{array}{l l} 2[1+\ln(8\dot{m}/25)], \ \ & \mbox{if}\, \
\dot{m}>25/8 \\ 
16\dot{m}/25, \ \ & \mbox{otherwise}.  
\end{array} \right.
\label{eq:fedd} 
\ee Note that the above formula looks different from that in the
reference paper, simply because the $\dot{M}_\edd$ defined here is
$16$ times larger than that in \citet{min00}.  Then according to
$\eta=f_\edd/(16\dot{m})$, we have \be
\eta=\left\{
\begin{array}{l l} [1+\ln(8\dot{m}/25)]/(8\dot{m}), \ \ & \mbox{if}\,
\ \dot{m}>25/8 \\
1/25, \ \ & \mbox{otherwise}.
\end{array} \right.
\label{eq:eta}
\ee For thick disc accretion, we have $R_\mb<R_\in<R_{\rm ISCO}$
\citep{koz78, Jaro80}, where
$R_\mb=2 - a+2(1 - a)^{1/2}$
%
is the marginally bound orbit.  In order to calculate the energy and
angular momentum brought into the central MBH through the inner boundary
of the accretion disc, we approximately estimate $R_\in$ according to
the following procedures. (i) For a given $\dot{m}$, $\eta$ is
estimated from Equation~(\ref{eq:eta}), and $E(R_\in)=1-\eta$ assuming
that the kinetic energy output is negligible.  (ii) According to
Equation~(\ref{eq:ein}), $E(r)$ can be obtained for any given radius
$r$ provided known spin $a$, and thus an array of $r$ and $E(r)$ can
be obtained. (iii) With the $E(R_\in)$ derived in step (i), $R_\in$ is
then obtained by interpolation of the array, and with
Eq.~(\ref{eq:phiin}) the specific angular momentum at $R_\in$ can be
obtained.

\section{properties of accretion disc in chaotic phase}
\label{sec:app3}

The disc in the chaotic phase is assumed to be described by the
thin disc model \citep[][]{sha73}, and the properties of the disc has
been studied in detail by \citet{per09} and \citet{dot13}. Here we
only give a brief summary of them. 

We assume a power-law profile of radial and vertical shear viscosity,
i.e., $\nu_1 \propto R^{3/4}$ and $\nu_2 \propto R^{3/4}$. The
$\alpha$-prescription is applied and $\alpha=0.09$ for radial
viscosity.  Choosing a different $\alpha$ does not significantly
affect our results.  For a disc with mass $M_\disc=M_\cl = b
M^{\gamma+1}_\bh$ (Eq.~\ref{eq:mcl}), the surface density profile is
given by the outer solution of the standard thin disc
model\footnote{Note that the solution to the standard thin disc
accretion at the inner region is different from that in the outer
region. In the cases of small discs, the disc solution to the outer
region may be not applicable. We have tested it by adopting the
solution to the inner region for those small discs and found that it
makes little difference to the MBH spin evolution.}, i.e., 
\be
\Sigma(R)=\Sigma_0(R/2 r_{\rm g})^{-3/4},
\ee 
where $r_{\rm g}=GM_\bh/c^2$ is the gravitational radius and 
\be
\Sigma_0 = 4 \times 10^7 \ \alpha^{-4/5}_{0.1} M^{1/5}_{\bh, 6}
\foe^{7/10}   {\rm g \ cm^{-2}}.
\label{eq:sigma}
\ee 
Here $\alpha_{0.1}=\alpha/0.1$, $\eta_{0.1}=\eta/0.1$, and
$M_{\bh,6}$ is MBH mass in unit of $10^6 M_\odot$. The disc size can
then be derived as 
\be
R_{\disc} \approx  && 10^5 \left(\frac{b}{0.01}\right)^{4/5}
\alpha^{16/25}_{0.1} M^{-24/25}_{\bh, 6}   \foe^{-14/25} r_{\rm g}.  
\label{eq:rbh}
\ee 
In the first order approximation, the disc angular momentum per
unit area is given by 
\be
L(R) \approx \frac{\dot{M}}{3\pi\nu_1}\sqrt{GM_\bh R}.
\ee 
Then the angular momentum of the whole disc can be derived by
integrating $L(R)$ over all annuli and set $R=R_\disc$, which yields
\be
J_\disc \propto \dot{M}\sqrt{GM_\bh} R^{7/4}_{\disc}. 
\ee 
The disc-to-MBH angular momentum ratio, which governs the MBH spin
evolution, can then be expressed as 
\be
\frac{J_\disc}{J_\bh} \approx 3\ \left(\frac{b}{0.01}\right)^{7/5}
\alpha^{8/25}_{0.1} M^{-12/25}_{\bh, 6} \foe^{-7/25} a^{-1}. 
\label{eq:jdojh}
\ee

The general misalignment between the angular momenta of disc and MBH
spin induces deformation in the disc \citep{bar75}, which is maximally
warped at around 
\be
R_\wp \approx 1000 \ \alpha^{24/35}_{0.1} f^{-4/7}_{\nu_2}
M^{4/35}_{\bh, 6} \foe^{-6/35} a^{4/7} r_{\rm g}, 
\label{eq:rwp}  
\ee  
where $f_{\nu_2}=2\alpha^2 \nu_2/\nu_1$ is a coefficient
describing non-linear effect ant $f_{\nu_2}= 0.6$ according to
simulation results \citep[see][for details]{lod07, per09}.  The
equality $R_\wp=R_\disc$ defines a critical MBH mass of 
\be
M^\crit_{\bh} \approx &&10^7 M_\odot \ \left(\frac{M_\disc}{10^4
M_\odot}\right)^{35/82} \alpha^{-1/41}_{0.1} a^{-25/82}  \foe^{-17/82}
f^{25/82}_{\nu_2}.
\label{eq:mcrit}
\ee

As for the self-gravitated disc, the criterion for disc stability is
given by the Toomre-$Q$ parameter with $Q\sim \Omega c_s /\pi G
\Sigma$ \citep[see][]{BT08}, where $\Omega$ is the angular velocity
and $c_s$ is the sound speed. The critical case of $Q=1$ yields a
maximum disc radius 
\be
R_{\disc,\sg} \approx 2 \times 10^5 \ \alpha^{28/45}_{0.1}
M^{-52/45}_{\bh,6} \foe^{-22/45} r_{\rm g}, 
\label{eq:rsg}
\ee 
which implies a possible upper limit on the disc mass, i.e., 
\be
M_\sg \approx 2 \times10^4  \ \alpha^{-1/45}_{0.1} \foe^{4/45}
M^{34/45}_{\bh,6} M_\sun.
\label{eq:msg}
\ee


\begin{thebibliography}{}

\bibitem[Abramowicz et al.(1988)]{abr88} Abramowicz, M. A., Czerny,
B., Lasota, J. P., \& Szuszkiewicz, E. 1988, \apj, 332, 646
%
\bibitem[Ag\'{i}s-Gonz\'{a}lez et al.(2014)] {ag14}
Ag\'{i}s-Gonz\'{a}lez, B., Miniutti, G., Kara, E., et al. 2014,
\mnras, 443, 2862 
%
\bibitem[Assef et al.(2011)] {assef11} Assef, R. J., Denney, K. D.,
Kochanek, C. S., et al. 2011, \apj, 742, 93
%
\bibitem[Aversa et al.(2015)]{Aversa15} Aversa, R., Lapi, A., de
Zotti, G., Shankar, F., \& Danese, L.\ 2015, \apj, 810, 74 
%
\bibitem[Ba{\~n}ados et al.(2018)]{Banados18} Ba{\~n}ados, E.,
Venemans, B.~P., Mazzucchelli, C., et al.\ 2018, \nat, 553, 473 
%
\bibitem[Bardeen \& Petterson(1975)] {bar75} Bardeen, J. M., \&
Petterson, J. A. 1975, \apj, 195, 65
%
\bibitem[Bardeen et al.(1972)] {bar72} Bardeen, J. M., Press, W. H.,
\& Teukolsky, S. A. 1972, \apj, 178, 347
%
\bibitem[Bennert et al.(2011)] {bennert11} Bennert, V. N., Auger, M.
W., Treu, T., Woo, J.-H., \& Malkan, M. A. 2011, \apj, 726, 59
%
\bibitem[Bentz et al.(2006)] {bentz06} Bentz, M. C., Denney, K. D.,
Cackett, E. M., et al. 2006, \apj, 651, 775
%
\bibitem[Bernardi et al.(2007)]{bernardi07} Bernardi, M., Sheth, R.
K., Tundo, E., \& Hyde, B. 2007, \apj, 660, 267
%
\bibitem[Berti \& Volonteri(2008)] {ber08} Berti, E., Volonteri, M.
2008, \apj, 684, 822
%
\bibitem[Binney \& Tremaine(2008)]{BT08} Binney, J., \& Tremaine, S.\
2008, Galactic Dynamics: Second Edition, by James Binney and Scott
Tremaine.~ISBN 978-0-691-13026-2 (HB).~Published by Princeton
University Press, Princeton, NJ USA, 2008
%
\bibitem[Brenneman(2013)] {bre13} Brenneman, L. 2013, AcPol, 53, 652
%
\bibitem[Brenneman \& Reynolds(2006)]{BR06} Brenneman, L.~W., \&
Reynolds, C.~S.\ 2006, \apj, 652, 1028 
%
\bibitem[Brenneman et al.(2011)]{brenneman11} Brenneman, L. W.,
Reynolds, C. S., Nowak, M. A., et al. 2011, \apj, 736, 103
%
\bibitem[Centrella et al.(2010)]{Centrella10} Centrella, J., Baker,
J.~G., Kelly, B.~J., \& van Meter, J.~R.\ 2010, Reviews of Modern
Physics, 82, 3069 
%
\bibitem[Czerny et al.(2011)]{czerny11} Czerny, B., Hryniewicz, K.,
Niko{\l}ajuk, M., \& Sadowski, A. 2011, \mnras, 415, 2942
%
\bibitem[Davis \& Laor(2011)]{DL11} Davis, S.~W., \& Laor, A.\ 2011,
\apj, 728, 98 
%
\bibitem[Dotti et al.(2013)] {dot13} Dotti, M., Colpi, M., Pallini,
S., et al. 2013, \apj, 762, 68
%
\bibitem[Du et al.(2015)]{Du15} Du, P., Hu, C., Lu, K.-X., et al.\
2015, \apj, 806, 22 
%
\bibitem[Dubois et al.(2014a)]{Dubois14_1} Dubois, Y., Volonteri, M.,
Silk, J., Devriendt, J., \& Slyz, A.\ 2014a, \mnras, 440, 2333 
%
\bibitem[Dubois et al.(2014b)]{Dubois14_2} Dubois, Y., Volonteri, M.,
\& Silk, J.\ 2014b, \mnras, 440, 1590 
%
\bibitem[Elvis et al.(2002)]{Elvis02} Elvis, M., Risaliti, G., \&
Zamorani, G.\ 2002, \apjl, 565, L75 
%
\bibitem[Fabian et al.(2013)]{fabian13} Fabian, A. C., Kara, E.,
Walton, D. J., et al. 2013, \mnras, 429, 2917
%
\bibitem[Fabian et al.(1989)] {fab89} Fabian, A. C., Rees, M. J.,
Stella, L., \& White, N. E. 1989, \mnras, 238, 729
%
\bibitem[Ferrarese et al.(2000)] {fer00} Ferrarese, L., \& Merritt, D.
2000, \apj, 539, L9
%
\bibitem[Fiacconi et al.(2018)]{2018MNRAS.477.3807F} Fiacconi, D., 
Sijacki, D., \& Pringle, J.~E.\ 2018, \mnras, 477, 3807 
%
\bibitem[Frederick et al.(2018)]{2018arXiv180206056F} Frederick,
S.~E., Kara, E., Reynolds, C.~S., Pinto, C., \& Fabian, A.~C.\ 2018,
\apj, 867, 67 
%
\bibitem[Gallo et al.(2011)]{gallo11} Gallo, L. C., Miniutti, G.,
Miller, J. M., et al. 2011, \mnras, 411, 607
%
\bibitem[Gallo et al.(2015)]{gallo15} Gallo, L. C., Wilkins, D. R.,
Bonson, K., et al. 2015, \mnras, 446, 633
%
\bibitem[Gammie et al.(2004)]{Gammie04} Gammie, C.~F., Shapiro, S.~L.,
\& McKinney, J.~C.\ 2004, \apj, 602, 312 
%
\bibitem[Gebhardt et al.(2000)] {geb00} Gebhardt, K., Bender, R.,
Bower, G., et al. 2000, \apj, 539, L13
%
\bibitem[Gebhardt \& Thomas(2009)] {geb09} Gebhardt, K., \& Thomas, J.
2009, \apj, 700, 1690
%
\bibitem[Goodman \& Tan(2004)]{goo04} Goodman, J., \& Tan, J. C.
2004, \apj, 608, 108
%
\bibitem[Gonz\'{a}lez-Mart\'{i}n \& Vaughan(2012)] {gs12}
Gonz\'{a}lez-Mart\'{i}n, O., Vaughan, S. 2012, \aap, 544, A80
%
\bibitem[G\"{u}ltekin et al.(2009)] {gul09} G\"{u}ltekin, K.,
Richstone, D. O., Gebhardt, K., et al. 2009, \apj, 698, 198
%
\bibitem[Hopkins \& Hernquist(2006)]{Hopkins06} Hopkins, P.~F., \&
Hernquist, L.\ 2006, \apjs, 166, 1 
%
\bibitem[Hopkins et al.(2007)]{hopkins07} Hopkins, P. F., Richards, G.
T., \& Hernquist, L. 2007, \apj, 654, 731
%
\bibitem[Jaroszynski et al.(1980)]{Jaro80} Jaroszynski, M.,
Abramowicz, M.~A., \& Paczynski, B.\ 1980, Acta Astronomica, 30, 1 
%
\bibitem[Jiang et al.(2014)]{jia14} Jiang, Y-F., Stone, J. M., \&
Davis, S. W. 2014, \apj, 796, 106
%
\bibitem[Keck et al.(2015)]{keck15} Keck, M. L., Brenneman, L. W.,
Ballantyne, D. R., et al. 2015, \apj, 806, 149
%
\bibitem[King et al.(2005)] {kin05} King, A. R., Lubow, S. H.,
Ogilvie, G. I., \& Pringle, J. E. 2005, \mnras, 363, 49
%
\bibitem[King \& Pringle(2006)]{kin06} King, A.~R., \& Pringle, J.~E.\
2006, \mnras, 373, L90 
%
\bibitem[King et al.(2008)]{King08} King, A.~R., Pringle, J.~E., \&
Hofmann, J.~A.\ 2008, \mnras, 385, 1621 
%
\bibitem[Kolykhalov \& Sunyaev(1980)]{kol80} Kolykhalov, P. I., \&
Sunyaev, R. A. 1980, SvAL, 6, 357
%
\bibitem[Kormendy \& Richstone(1995)] {kor95} Kormendy, J., \&
Richstone, D. 1995, \araa, 33, 581
%
\bibitem[Kormendy \& Ho(2013)] {kor13} Kormendy, J., \& Ho, L. C.
2013, \araa, 51, 511
%
\bibitem[Koz\l owski et al.(1978)] {koz78} Koz\l owski, M.,
Jaroszy\'{n}ski, M., \& Abramowicz, M.~A. 1978, \aap, 63, 209
%
\bibitem[Krolik(1999)]{Krolik99} Krolik, J.~H.\ 1999, Active galactic
nuclei : from the central black hole to the galactic environment
/Julian H.~Krolik.~Princeton, N.~J.~: Princeton University Press,
c1999.,  
%
\bibitem[LaMassa et al.(2015)]{LaM15} LaMassa, S.~M., Cales, S.,
Moran, E.~C., et al.\ 2015, \apj, 800, 144 
%
\bibitem[Laor(1991)] {lao91} Laor, A. 1991, \apj, 376, 90
%
\bibitem[Lapi et al.(2006)]{Lapi06} Lapi, A., Shankar, F., Mao, J., et
al.\ 2006, \apj, 650, 42 
%
\bibitem[Lehner \& Pretorius(2014)]{Lehner14} Lehner, L., \&
Pretorius, F.\ 2014, \araa, 52, 661 
%
\bibitem[Lense \& Thirring(1918)] {len18} Lense, J., \& Thirring, H.
1918, Phys. Z., 19, 156
%
\bibitem[Levinson \& Nakar(2018)] {lev18} Levinson, A., \& Nakar, E.
2018, \mnras, 473, 2673
%
\bibitem[Li(2012)]{Li12} Li, L.-X.\ 2012, \mnras, 424, 1461 
%
\bibitem[Li et al.(2015)] {li15} Li, Y.-R., Wang, J.-M., Cheng, C., \&
Qiu, J. 2015, \apj, 804, 45
%
\bibitem[Lodato et al.(2006)]{lodato06} Lodata, G., \& Pringle, J. E.
2006, \mnras, 368, 1196
%
\bibitem[Lodato \& Pringle(2007)] {lod07} Lodato, G., \& Pringle, J.
E. 2007, \mnras, 381, 1287
%
\bibitem[Lohfink et al.(2013)]{lohfink13} Lohfink, A. M., Reynolds, C.
S., Jorstad, S. G., et al. 2013, \apj, 772, 83
%
\bibitem[Lohfink et al.(2012)]{lohfink12} Lohfink, A. M., Reynolds, C.
S., Miller, J. M., et al. 2012, \apj, 758, 67
%
\bibitem[Lousto et al.(2010)]{Lousto10} Lousto, C.~O., Campanelli, M.,
Zlochower, Y., \& Nakano, H.\ 2010, Classical and Quantum Gravity, 27,
114006 
%
\bibitem[Macchetto et al.(1997)] {mac97} Macchetto, F., Marconi, A.,
Axon, D. J., et al. 1997, \apj, 489, 579 
%
\bibitem[Madau et al.(2014)] {mad14} Madau, P., Haardt, F., \& Dotti,
M. 2014, \apj, 784, 38
%
\bibitem[Magorrian et al.(1998)] {mag98} Magorrian, J., Tremaine, S.,
Richstone, D., et al. 1998, \apj, 115, 2285
%
\bibitem[Malizia et al.(2008)]{malizia08} Malizia, A., Bassani, L.,
Bird, A. J., et al.  2008, \mnras, 389, 1360
%
\bibitem[Marconi et al.(2004)]{Marconi04} Marconi, A., Risaliti, G.,
Gilli, R., et al.\ 2004, \mnras, 351, 169 
%
\bibitem[Martini(2004)]{Martini04} Martini, P.\ 2004, Coevolution of
Black Holes and Galaxies, 169 
%
\bibitem[Martine et al.(2007)] {mar07} Martin, R. G., Pringle, J. E.,
\& Tout, C. A. 2007, \mnras, 381, 1617
%
\bibitem[McHardy et al.(2005)] {mchardy05} McHardy, I. M., Gunn, K.
F., Uttley, P., \& Goad, M. R. 2005, \mnras, 359, 1469
%
\bibitem[Mineshige et al.(2000)] {min00} Mineshige, S., Kawaguchi, T.,
Takeuchi, M., \& Hayashida, K. 2000, PASJ, 52, 499
%
\bibitem[Miniutti et al.(2009)]{miniutti09} Miniutti, G., Panessa, F.,
De Rosa, A., et al. 2009, \mnras, 398, 255
%
\bibitem[Miniutti et al.(2010)] {miniutti10} Miniutti, G., Piconcelli,
E., Bianchi, S., Vignali, C., \& Bozzo, E. 2010, \mnras, 401, 1315
%
\bibitem[Moderski et al.(1998)]{Moderski98} Moderski, R., Sikora, M.,
\& Lasota, J.-P.\ 1998, \mnras, 301, 142 
%
\bibitem[Mortlock et al.(2011)] {mor11} Mortlock, D. J., Warren, S.
J., Venemans, B. P., et al. 2011, \nat, 474, 616
%
\bibitem[Niko\l ajuk et al.(2009)] {nikolajuk09} Niko\l ajuk, M.,
Czerny, B., \& Gurynowicz, P. 2009, \mnras, 394, 2141
%
\bibitem[Novikov \& Thorne(1973)]{NT73} Novikov, I.~D., \& Thorne,
K.~S.\ 1973, Black Holes (Les Astres Occlus), 343 
%
\bibitem[Patrick et al.(2012)]{patrick12} Patrick, A. R., Reeves, J.
N., Porquet, D., et al. 2012, \mnras, 426, 2522
%
\bibitem[Perego et al.(2009)]{per09} Perego, A., Dotti, M., Colpi, M.,
\& Volonteri, M. 2009, \mnras, 399, 2249
%
\bibitem[Peterson et al.(2004)] {pet04} Peterson, B. M., Ferrarese,
L., Gilbert, K. M., et al. 2004, \apj, 613, 682 
%
\bibitem[Press et al.(2007)]{pre07} Press, W. H., Teukolsky, S. A.,
Vetterling, W. T., \& Flannery, B. P. 2007, Numerical recipes, 3rd
edn.  Cambridge Univ. Press, Cambridge, p. 762
%
\bibitem[Raimundo \& Fabian(2009)]{Raimundo09} Raimundo, S.~I., \&
Fabian, A.~C.\ 2009, \mnras, 396, 1217
%
\bibitem[Raimundo et al.(2012)]{Raimundo12} Raimundo, S.~I., Fabian,
A.~C., Vasudevan, R.~V., Gandhi, P., \& Wu, J.\ 2012, \mnras, 419,
2529
%
\bibitem[Reis et al.(2014)]{Reis14} Reis, R.~C., Reynolds, M.~T.,
Miller, J.~M., \& Walton, D.~J.\ 2014, \nat, 507, 207 
%
\bibitem[Reynolds et al.(2014)] {reynolds14} Reynolds, C. S., Lohfink,
A. M., Babul, A., et al.  2014, \apj, 792, L41
%
\bibitem[Reynolds(2014)]{rey14} Reynolds, C. S. 2014, SSRv, 183, 277
%
\bibitem[Reynolds et al.(2014)] {rey_mt14} Reynolds, M. T., Walton, D.
J., Miller, J. M., \& Reis, R. 2014, \apjl, 792, L19
%
\bibitem[Risaliti et al.(2013)] {risaliti13} Risaliti, G., Harrison,
F. A., Madsen, K. K., et al.  2013, \nat, 494, 449
%
\bibitem[Risaliti et al.(2009)]{risaliti09} Risaliti, G., Miniutti,
G., Elvis, M., et al.  2009, \apj, 696, 160
%
\bibitem[Ruan et al.(2016)]{Ruan16} Ruan, J.~J., Anderson, S.~F.,
Cales, S.~L., et al.\ 2016, \apj, 826, 188 
%
\bibitem[S\k{a}dowski(2009)] {sad09} S\k{a}dowski, A. 2009, ApJS, 183,
171 
%
\bibitem[S\k{a}dowski et al.(2011)] {sad11} S\k{a}dowski, A., Bursa,
M., Abramowicz, M., et al. 2011, \aap, 532, 41
%
\bibitem[Saglia et al.(2016)] {sag16} Saglia, R. P., Opitsch, M.,
Erwin, P., et al. 2016, \apj, 818, 47
%
\bibitem[Schulze \& Wisotzki(2010)] {sch10} Schulze, A., \& Wisotzki,
L. 2010, \aap, 516, A87
%
\bibitem[Sesana et al.(2014)] {ses14} Sesana, A., Barausse, E., Dotti,
M., Rossi, E. M. 2014, \apj, 794, 104
%
\bibitem[Shakura \& Sunyaev(1973)] {sha73} Shakura, N. I., \& Sunyaev,
R. A. 1973, \aap, 24, 337
%
\bibitem[Shankar et al.(2016)]{shankar16} Shankar, F., Bernardi, M.,
Sheth, R. K. 2016, \mnras, 460, 3119
%
\bibitem[Shankar et al.(2004)]{Shankar04} Shankar, F., Salucci, P.,
Granato, G.~L., De Zotti, G., \& Danese, L.\ 2004, \mnras, 354, 1020 
%
\bibitem[Shankar et al.(2009)]{Shankar09} Shankar, F., Weinberg,
D.~H., \& Miralda-Escud{\'e}, J.\ 2009, \apj, 690, 20
%
\bibitem[Shankar et al.(2013)]{shankar13} Shankar, F., Weinberg,
D.~H., \& Miralda-Escud{\'e}, J.\ 2009, \mnras, 428, 421
%
\bibitem[Shen et al.(2008)]{she08} Shen, Y., Greene, J. E., Strauss,
M. A., Richards, G. T., \& Schneider, D. P. 2008, \apj, 680, 169
%
\bibitem[Shen(2009)]{Shen09} Shen, Y.\ 2009, \apj, 704, 89 
%
\bibitem[Shen et al.(2011)]{she11} Shen, Y., Richards, G. T., Strauss,
M. A., et al. 2011, \apjs, 194, 45
%
\bibitem[Sluse et al.(2012)] {sluse12} Sluse, D., Hutsem\'{e}kers, D.,
Courbin, F., Meylan, G., \& Wambsganss, J. 2012, \aap, 544, A62
%
\bibitem[So{\l}tan(1982)]{1982MNRAS.200..115S} So{\l}tan, A.\ 1982,
\mnras, 200, 115 
%
\bibitem[Suh et al.(2015)]{Suh15} Suh, H., Hasinger, G., Steinhardt,
C., Silverman, J.~D., \& Schramm, M.\ 2015, \apj, 815, 129 
%
\bibitem[Sun et al.(2017)]{sun17} Sun, S., Guainazzi, M., Ni, Q., et
al.  2017, arXiv:1704.03716
%
\bibitem[Tanaka et al.(1995)]{Tanaka95} Tanaka, Y., Nandra, K.,
Fabian, A.~C., et al.\ 1995, \nat, 375, 659 
%
\bibitem[Tan et al.(2012)]{tan12} Tan, Y., Wang, J. X., Shu, X. W., \&
Zhou, Y.  2012, \apj, 747, L11
%
\bibitem[Thorne(1974)] {tho74} Thorne, K. S. 1974, \apj, 191, 507
%
\bibitem[Tremaine et al.(2002)] {tre02} Tremaine, S., Gebhardt, K.,
Bender, R., et al. 2002, \apj, 574, 740
%
\bibitem[Ueda et al.(2014)]{ueda14} Ueda, Y., Akiyama, M., Hasinger,
G., et al. 2014, \apj, 786, 104
%
\bibitem[Vasudevan et al.(2016)] {vas16} Vasudevan, R. V., Fabian, A.
C., Reynolds, C. S., et al. 2016, \mnras, 458, 2012
%
\bibitem[Vestergaard \& Peterson(2006)]{Vestergaard06} Vestergaard,
M., \& Peterson, B.~M.\ 2006, \apj, 641, 689 
%
\bibitem[Volonteri et al.(2013)]{Volonteri13} Volonteri, M., Sikora,
M., Lasota, J.-P., \& Merloni, A.\ 2013, \apj, 775, 94 
%
\bibitem[Volonteri et al.(2005)] {vol05} Volonteri, M., Madau, P.,
Quataert, E., \& Rees, M. J. 2005, \apj, 620, 69
%
\bibitem[Walton et al.(2013)]{walton13} Walton, D. J., Nardini, E.,
Fabian, A. C., Gallo, L. C., \& Reis, R. C. 2013, \mnras, 428, 2901
%
\bibitem[Wu et al.(2013)]{Wu13} Wu, S., Lu, Y., Zhang, F., \& Lu, Y.\
2013, \mnras, 436, 3271 
%
\bibitem[Wu et al.(2015)] {wu15} Wu, X-B., Wang, F., Fan, X., et al.
2015, \nat, 518, 512
%
\bibitem[Yu \& Lu(2004)]{YL04} Yu, Q., \& Lu, Y.\ 2004, \apj, 602, 603 
\bibitem[Yu \& Lu(2008)]{yu08} Yu, Q., \& Lu, Y. 2008, \apj, 689, 732
%
\bibitem[Yu et al.(2005)]{YLK05} Yu, Q., Lu, Y., \& Kauffmann, G.\
2005, \apj, 634, 901 
%
\bibitem[Yu \& Tremaine(2002)] {yu02} Yu, Q., \& Tremaine, S. 2002,
\mnras, 335, 965
%
\bibitem[Zhang \& Lu(2017)]{ZL17} Zhang, X., \& Lu, Y.\ 2017, Science
China Physics, Mechanics, and Astronomy, 60, 109511 
%
\bibitem[Zhang et al.(2012)] {zha12} Zhang, X., Lu, Y., \& Yu, Q.
2012, \apj, 761, 5
%
\bibitem[Zhang et al.(2018)]{ZLZ18} Zhang, X., Lu, Y., \& Liu, Z.\
2018, submitted to ApJ
%
\bibitem[Zhou \& Wang(2005)] {zhou05} Zhou, X.-L., \& Wang, J.-M.
2005, \apj, 618, L83
%
\bibitem[Zoghbi et al.(2010)] {zoghbi10} Zoghbi, A., Fabian, A. C.,
Uttley, P., et al.  2010, \mnras, 401, 2419


\end{thebibliography}
\end{document}